%% file: main.tex
\newif\ifTwoColumn
\TwoColumntrue            

\ifTwoColumn
\documentclass[nofootinbib,groupedaddress,pra,aps,10pt,twocolumn,floatfix]{revtex4-2}
\else
\documentclass[nofootinbib,groupedaddress,pra,aps,10pt,onecolumn]{revtex4-2}
\fi
\pdfoutput=1

\input{preamble}            

\makeatletter
\clo@groupedaddress
\makeatother

\begin{document}

\title{High-Accuracy Propagation-Assisted Local Decoding for Fast Transversal Clifford Circuits}

\author{Mark Webber\textsuperscript{*\,\textdagger}\, \orcidlink{0000-0002-8674-9439}}
\author{Simanraj Sadana\textsuperscript{*}\, \orcidlink{0000-0003-0561-4442}}
\author{Matthew A. Lane \orcidlink{0000-0002-3222-1348}}
\author{Lucy Jade Robson \orcidlink{0009-0008-2440-7960}}
\author{Anthony O'Rourke \orcidlink{0009-0000-4736-7537}}

\affiliation{Universal Quantum Ltd, Gemini House, Haywards Heath, United Kingdom}

\date{\today}

\begin{abstract}
We benchmark the Iterative Redecode Scheme (IRS) for fast transversal Clifford circuits and show that high-accuracy correlated decoding is possible using local, memory-style detectors and decoders. On a deep random Clifford benchmark, IRS achieves approximately $4\times$ lower logical failure rates than the ghost protocol~\cite{turner2026scalable} at $d=11$ with the same syndrome-extraction buffer, and $15\times$ lower rates than logical observable matching (LOM)~\cite{serra2026decoding}. In three-block GHZ preparation, IRS achieves $2\times$ lower logical error rates than reliable logical Pauli-product decoding~\cite{cain2025fast} with matched initialisation, improving to $25\times$ at $d=19$ with  $\lceil d/2\rceil$ full initialisation rounds. We build on the iterative scheme of Wan \emph{et al.}~\cite{wan2024iterative} within a broader propagation-assisted local (PAL) framework, in which cross-block correlations are communicated through messages rather than decoded as a global hypergraph. IRS alternates local memory decoding, here using minimum-weight perfect matching (MWPM) against memory-equivalent detector error models (DEMs), with Pauli-frame propagation and detector updates through successive transversal CNOT (tCNOT) layers, seeking local error configurations that, together with their propagated contributions, explain the measured global detector record. We introduce a prior-informed DEM that accounts for the probability of propagated errors absent from the memory-equivalent model, improving initial decoding and reducing logical error rates. We also introduce zero-weight Pauli mechanisms to infer and propagate initially indeterminate stabilizer outcomes. Each IRS iteration defines a deterministic map on a finite state space, so the uncapped procedure must reach a fixed point or a cycle detectable through state recurrence. We show that even on deep, adversarial alternating-tCNOT circuits, logical error rates saturate after only a few decoding iterations, allowing the iteration count to be capped with negligible loss of accuracy. We also investigate syndrome-extraction buffers, finding separations of order $\lfloor d/4\rfloor$ sufficient to maintain distance-dependent error suppression in the demanding regimes studied. IRS is not restricted to MWPM or the surface code: the same principles extend to other local decoders and CSS codes. We also demonstrate applicability beyond tCNOTs by decoding repeated logical $S$ gates in the surface code with approximately $4$--$10\times$ lower logical error rates than LOM~\cite{serra2026decoding}, depending on the initialisation schedule.

\end{abstract}
\makeatletter
\g@addto@macro\@FMN@list{%
  \@TBN@opr{1}{These authors contributed equally to the technical work of this manuscript.}%
  \@TBN@opr{2}{Corresponding author: \href{mailto:m.webber@universalquantum.com}{m.webber@universalquantum.com}}%
}
\makeatother

\maketitle

\section{Introduction}\label{sec:introduction}
\subsection{Problem Statement}
Fault-tolerant quantum computing using the surface code has historically focused on lattice-surgery and braid-based logical operations, reflecting the assumption of nearest-neighbour connectivity in two-dimensional architectures \cite{dennis2002topological,fowler2012surface,horsman2012surface,litinski2019game,fowler2018low}.
These constructions are well matched to planar superconducting architectures and admit efficient graph-based decoding methods based on the local detector structure \cite{delfosse2021almost,higgott2025sparse,google2025quantum}.
However, several hardware platforms naturally support long-range physical connectivity \cite{cohen2022low,xu2024constant}, including trapped-ion systems through ion transport and reconfigurable neutral-atom arrays through dynamic rearrangement \cite{ryan2021realization,bluvstein2022quantum,bluvstein2024logical}.

In such architectures, some logical Clifford operations can instead be implemented as \textit{transversal} gates, which can be performed with a circuit that has constant depth with respect to code distance $d$.
A partition of the qubits of a code is said to be transversal if it only contains one qubit from each code block~\cite{eastin2009restrictions}.
A transversal logical gate on a single code block is thus a tensor product of single-qubit gates, while on multiple code blocks comprises multi-qubit physical gates that couple disjoint sets of qubits that contain only one qubit per block~\cite{eastin2009restrictions}.
There also exist ``fold''-transversal gates that pair qubits in a single code block.
In the surface code they are a tensor product of single-qubit gates along a central diagonal, or ``fold'', and two-qubit gates pairing qubits with mirrored positions about this fold~\cite{moussa2016transversal}.
This is generalised to all CSS codes in \cite{breuckmann2024fold}, as being a tensor product of single qubit unitaries, and two-qubit unitaries on qubits that are in the same orbit of a $ZX$-duality.
For Calderbank-Shor-Steane (CSS) codes, there is a transversal CNOT (tCNOT), giving constant-depth logical entangling operations without the $O(d)$ syndrome-extraction overhead associated with lattice surgery \cite{steane1997active,knill2004fault,kubica2015unfolding,moussa2016transversal,breuckmann2024fold,chen2026transversal}.
A tCNOT requires only a single layer of pairwise physical CNOT operations between logical blocks. 
However, continuing to adequately suppress error rates in circuits with successive transversal gates can depend on the circuit structure, the physical error model, code-distance, and the decoder.
We primarily study the low-overhead regime of a single intermediate syndrome-extraction round between layers of tCNOTs, while also investigating when additional rounds improve logical performance, at the cost of increased circuit time complexity.
The resulting reduction in the depth of the logical circuit is particularly valuable for Clifford-dominated fault-tolerant primitives \cite{litinski2019game,beverland2021cost,bluvstein2024logical}, including magic-state distillation, where large layers of logical CNOT operations contribute substantially to both depth and resource cost \cite{bravyi2005universal,gidney2019efficient,sales2025experimental}.

These benefits come at the cost of a harder decoding problem.
For transversal gates built from entangling physical interactions, such as the tCNOT, and certain implementations of the logical $S$ gate, faults can propagate and generate a correlated detector structure across and within logical blocks.
In contrast, transversal operations composed only of single-qubit physical gates, such as logical Pauli operations and the Hadamard implemented with edge-tracking~\cite{litinski2019game}, do not spread errors between physical qubits and do not give rise to propagation-induced correlations.

We take the tCNOT as the canonical example of the non-local detector correlations generated when transversal Clifford operations couple logical blocks and propagate Pauli errors between them. Under Clifford evolution, Pauli errors are deterministically transported through the logical circuit, so a single physical fault near a transversal gate can generate correlated detector events across multiple blocks and syndrome-extraction rounds \cite{moussa2016transversal,cain2024correlated,sahay2025error,turner2026scalable,serra2026decoding}. For a tCNOT, these propagated faults produce higher-order detector correlations, in which one fault flips more than two detectors and cannot be decomposed into pairwise detector edges already present in the decoding graph \cite{cain2024correlated,sahay2025error,turner2026scalable,serra2026decoding}.

The resulting decoding hypergraph differs substantially from the local pairwise correlations assumed by standard MWPM decoders \cite{kolmogorov2009blossom,wang2011surface,higgott2022pymatching,higgott2025sparse}, and decoding each logical block independently would discard the information that several detector events may originate from the same propagated fault mechanism.
Decoding strategies developed for memory experiments and lattice-surgery circuits therefore do not always generalise directly to deep tCNOT circuits \cite{skoric2023parallel,tan2023scalable}.

\subsection{Related Work}\label{sec:prior-work}
Several approaches have been developed to address these propagated detector correlations:

\begin{itemize}[leftmargin=*]
    \item A general framework for correlated decoding of transversal logical circuits was introduced in \cite{cain2024correlated}. It represents the joint decoding problem as a cross-logical-qubit hypergraph and studies both exact Maximum Likelihood Estimation (MLE) and a more scalable belief propagation (BP) hypergraph-union-find (HUF) hybrid decoder. Their proposed BP-HUF decoder provides a tractable approximate approach with a significant performance gap relative to MLE in the low-logical-error regime, indicative of the difficulty of efficiently decoding the full correlated hypergraph.
    \item Refs.~\cite{wan2024iterative, wan2024constant} introduced, what we identify as, the first Propagation Assisted Local (PAL) decoder for tCNOT circuits. We use PAL to denote a broad class of methods where correlations between otherwise local decoding problems are passed between logical blocks as messages in the form of propagated detector updates, Pauli-frame information, or related metadata, rather than decoded as a single global hypergraph. The purpose of this message passing is to leave each block with a memory(graph)-like decoding subproblem, amenable to fast decoders such as MWPM. Rather than constructing and decoding the full correlated hypergraph, Ref.~\cite{wan2024iterative} retains local graph-like decoding problems and communicates the missing correlation information through explicit propagation of inferred error and Pauli-frame information, repeatedly re-decoding until the propagated interpretation becomes self-consistent, and is the starting point for the method developed in this work.
    \item Ref.~\cite{sahay2025error} compared several tCNOT decoding strategies, including HUF and their ordered-MWPM decoder based on exploiting deterministic error propagation between local decoding problems. Ordered decoding first decodes the logical qubit from which a relevant Pauli error propagates, applies the inferred propagated correction to the receiving logical qubit, and then decodes the residual syndrome there. Ordered decoding implements a single directed PAL update for the tCNOT, but lacks an iterative mechanism for reconciling mutually dependent local decoding solutions in more general circuits.
    \item Local propagation-assisted architecture was then developed in the \textit{ghost protocol}~\cite{turner2026scalable}, where cross-logical-qubit hyperedges are decomposed into local graph-like \emph{ghost} fragments, linked by metadata identifying the common parent hyperedge from which they originate. This allows each logical qubit to be decoded locally; committing to such a ghost fragment communicates the corresponding correlated update to the other local decoders through iterative message passing. This is an example of a PAL decoder.
    \item By contrast, recent approaches reformulate the decoding problem around matchable subgraphs of so-called reliable logical observables, enabling very efficient graph-like decoding \cite{serra2026decoding,cain2025fast} without the same propagate-and-redecode architecture. We refer to these schemes as LOM (Logical Observable Matching) decoders in general, as in \cite{serra2026decoding}, or as Reliable Logical Pauli-Products (RLPP) when referring specifically to the results of \cite{cain2025fast}. 
    \item Google Quantum AI's Tesseract decoder \cite{beni2025tesseract,grbic2026accelerating} uses an $A^*$ search over error configurations to approximate MLE decoding. It achieves high accuracy, including on tCNOT surface-code circuits, but published benchmarks show sharply increasing runtimes with code distance and physical error rate. Limiting the computational cost through search cutoffs can compromise decoding accuracy. Despite recent implementation improvements, the reported runtime scaling suggests that Tesseract may not be well suited to large-scale real-time decoding \cite{grbic2026accelerating}.
\end{itemize}

\subsection{Overview of this Work}\label{sec:paper-overview}
In this work, we develop PAL decoding as a general principle for correlated decoding problems resulting from error propagation, and introduce the Iterative Redecode Scheme (IRS) as a concrete implementation of it. Ref.~\cite{wan2024iterative} described its core procedure operationally in terms of iterative Pauli-frame message passing and syndrome modification, without identifying the broader decoding architecture that it represents.
Specifically, IRS alternates local decoding with propagation of inferred Pauli errors through successive tCNOT layers, storing the propagated contributions separately and cancelling their predicted detector signatures from the working record. Repeated decoding passes refine both the local error estimates and propagation corrections until they are consistent with the global-circuit detector record.
To this we add a propagation-informed decoding prior, a robust treatment of initialisation effects, and an analysis of convergence and cycling in deep transversal Clifford circuits. 

Nothing in the IRS construction is specific to MWPM, the surface code, or the tCNOT: IRS can be used with any existing local decoder as long as the decoder interface of the code satisfies two properties - a local memory decoding model exists, and an estimate of the Pauli error present at each gate boundary can be obtained - and applies to any logical gadget whose non-locality originates from internal entangling operations, including the fault-tolerant implementation of the $S$ gate considered in Section \ref{sec:S-gate}.
For tCNOTs, it can also be applied to any CSS code, provided that the two blocks are identical and their physical qubits are paired injectively, since a CSS code automatically admits a tCNOT with the required action on the code's stabilizers \cite{gottesman2024surviving}.
Here we construct and test this machinery for the surface code using MWPM as the local decoder, applied to the tCNOT and $S$ gate. 

We compare IRS directly with recent scalable transversal-decoding approaches across several established benchmarks, and find that it consistently achieves the lowest logical error rates in these comparisons, while retaining ordinary memory-style detector definitions and local decoding, with each block decoded using the model obtained by removing the tCNOTs and treating it as an independent memory.

Section \ref{sec:high-level-results} presents the headline result: a like-for-like comparison in which IRS outperforms the existing scalable methods outlined in Section \ref{sec:prior-work}. 
We then provide a concise overview of the methodology, before using each subsequent section to explore one component of the method in turn, and report results that quantify its contribution and compare it against the literature. 
Section \ref{sec:initial-stabilizers} describes our treatment of stabilizers that are indeterminate at initialisation. 
Section \ref{sec:iteration-dependence} examines how performance depends on iteration count and analyses the worst-case performance on pathological circuits that can prevent convergence.
Section \ref{sec:calibration} describes the construction of a prior-informed DEM and analyses the time-complexity of syndrome-extraction in the worst-case setting of alternating tCNOTs.
Section \ref{sec:S-gate} extends the method to an implementation of the $S$ gate.
Section \ref{sec:discussion} then discusses what these results imply for decoding fast transversal Clifford circuits, and for PAL decoding more broadly.
Much of the pedagogical detail and the supporting derivations are collected in the appendices. The Stim circuits and numerical data for all benchmarks are available in the accompanying Zenodo repository~\cite{webber2026stim}.

\section{High-Level Results and Methodology}\label{sec:high-level-results}
\subsection{High-Level Results}\label{sec:headline-results}
We study fast transversal logic, where the number of syndrome-extraction (SE) buffer rounds between logical-gate layers is small relative to the code distance $d$, and often $O(1)$ depending on circuit structure. For the adversarial circuits studied, we empirically find that $\lfloor d/4\rfloor$ buffer rounds suffice to maintain distance-dependent logical-error suppression with IRS. To benchmark IRS against other transversal decoders, we use the same fixed random circuit and noise model as Ref.~\cite{turner2026scalable}, comprised of Hadamard, Pauli operations and tCNOT layers, encoded in the rotated surface code; the error-model details are given in Appendix~\ref{app:noise-models}. The reference data for the ghost protocol~\cite{turner2026scalable} was supplied by its authors, and we used the open-source implementation of the LOM decoder~\cite{SerraPeralta_LOM,SerraPeralta_circuits}, validating our results against the published benchmark trends up to $d=11$.

In the deep random Clifford benchmark in Figure~\ref{fig:deep-random}, IRS achieves the lowest logical error rate per gate layer at every code distance sampled. With a single SE buffer round between logical-gate layers, its failure probability is lower than that of the ghost protocol~\cite{turner2026scalable} by a factor of $\sim 5$ at $d=9$ and $\sim 4$ at $d=11$. At $d=11$, it is also a factor of $\sim 15$ lower than logical observable matching (LOM)~\cite{serra2026decoding}. A second SE buffer round per layer can further improve all methods at $d=11$ (dashed lines), reducing the logical error rate per gate layer at the cost of a greater-depth syndrome-extraction schedule.

\begin{columnfigure}[htbp]
    \centering
    \includegraphics[width=\linewidth]{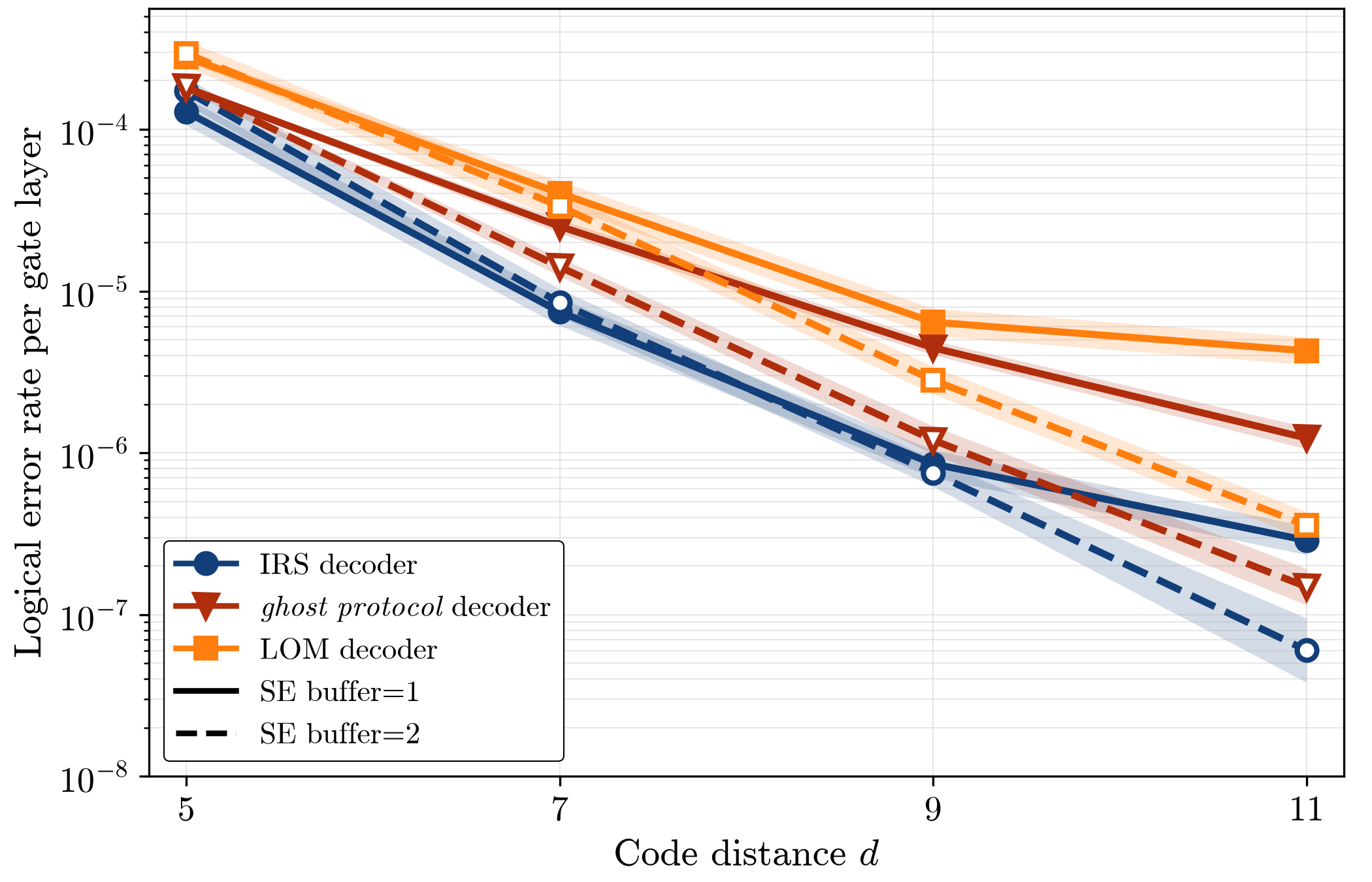}
    \caption{Deep random Clifford benchmark on the rotated surface code. Using the logical circuit and noise model of Ref.~\cite{turner2026scalable}, we plot the logical error rate per logical-gate layer versus code distance $d$ for: IRS \textit{(blue circles)} with a prior-informed initial pass and one or two syndrome-extraction (SE) buffer rounds between logical-gate layers, shown as solid and dashed lines, respectively; the ghost protocol \textit{(burgundy triangles)}~\cite{turner2026scalable} with one SE buffer round, using author-supplied reference data; and the open-source LOM  decoder \textit{(orange squares)}~\cite{serra2026decoding}. Our simulations accumulate approximately $100$ logical failures per point. IRS continues to suppress logical errors with increasing distance, with the benefit of a second SE buffer round emerging at $d=11$. Shaded bands show 95\% confidence Wilson intervals.}
    \label{fig:deep-random}
\end{columnfigure}

For the IRS decoder and challenging circuits such as the deep Clifford benchmark, our results suggest that an SE buffer of approximately $\lfloor d/4\rfloor$ rounds can maintain logical-error suppression with increasing code distance, avoiding the eventual plateau that may arise with a fixed one-round buffer. Each additional round supplies more information for decoding at the cost of introducing more noisy operations and increasing overall circuit depth and runtime, with the optimal separation being a function of circuit structure and time complexity. We examine this dependence more closely in Section~\ref{sec:calibration}, and the associated iteration behaviour in Section~\ref{sec:iteration-dependence}.

Ambiguity in the temporal matching of inferred errors around transversal gates is known to contribute to the distance dependence of local decoder performance. Specifically, closely spaced alternating tCNOTs can propagate an error from one block to another and then back to its source, causing the local matching solutions on the two blocks to become interdependent. This feedback motivates the iterative propagation-and-redecode procedure of IRS, rather than relying on a single pass. The inferred placement of errors immediately before or after a tCNOT is therefore crucial for determining their subsequent propagation, since errors inferred to occur after a tCNOT are not propagated by it. Increasing the separation between gates helps resolve these temporal ambiguities by providing additional syndrome information on either side of the gate. As code distance increases and other failure contributions are suppressed, ambiguities involving longer space-time error chains may become increasingly important in determining decoder performance and the optimal separation between tCNOTs. With a single-round buffer, the narrowing of IRS's advantage over ghost decoding from $\sim 5\times$ at $d=9$ to $\sim 4\times$ at $d=11$ may reflect the onset of a shared limitation associated with these temporal ambiguities, consistent with the improvement obtained by adding a second round at $d=11$.

Ref.~\cite{serra2026decoding} also improves LOM by applying belief propagation (BP) to the full decoding hypergraph before matching and using the resulting probabilities to incorporate correlation information into the LOM decoding graphs. This substantially improves LOM alone, but the reported logical error rates \cite{serra2026decoding} remain worse than those achieved by IRS on this benchmark, even with as many as $100$ BP iterations. The improvement also carries a substantial computational cost: BP radically dominates the cost of BP-assisted LOM. Plain LOM therefore provides a lightweight decoder with accuracy limitations for certain combinations of circuits and code distances.

The headline result of the IRS method is therefore high decoding accuracy with every local decoding call retaining the memory-style detector structure and matching machinery of an ordinary quantum memory.

\subsection{Core Methodology}\label{sec:methodology}
This section gives a working overview of the IRS decoding method, sufficient to read the detailed results that follow. The formal development, including the stabilizer-transport identities, the detector-record transformation, and the full algorithm, is given in the Appendices.

A circuit of tCNOTs comprises logical blocks coupled by transversal layers of physical CNOTs (see Figure~\ref{fig:tCNOT-gate-implementation}). Without these couplings, each block would be an ordinary memory experiment. IRS decodes each block against its \emph{memory-equivalent DEM}, obtained by removing the tCNOTs and treating the block as an independent memory. The detector definitions also remain local to each block, rather than incorporating cross-block stabiliser transport, so the local decoder never encounters a cross-block hyperedge. Instead, the consequences of transversal error propagation are handled by an outer classical layer which passes inferred information between blocks and modifies the detector records presented to the local decoders accordingly, with each decoding subproblem remaining local, while the propagation couples their solutions. The core aim of IRS is to find a self-consistent interpretation in which the local decoders explain the remaining detector events after inferred propagation contributions have been removed, and the resulting local error estimates, when combined with their propagated consequences, explain the original global detector record. 

\begin{columnfigure}[htbp]
    \centering
    \includegraphics[width=.9\linewidth]{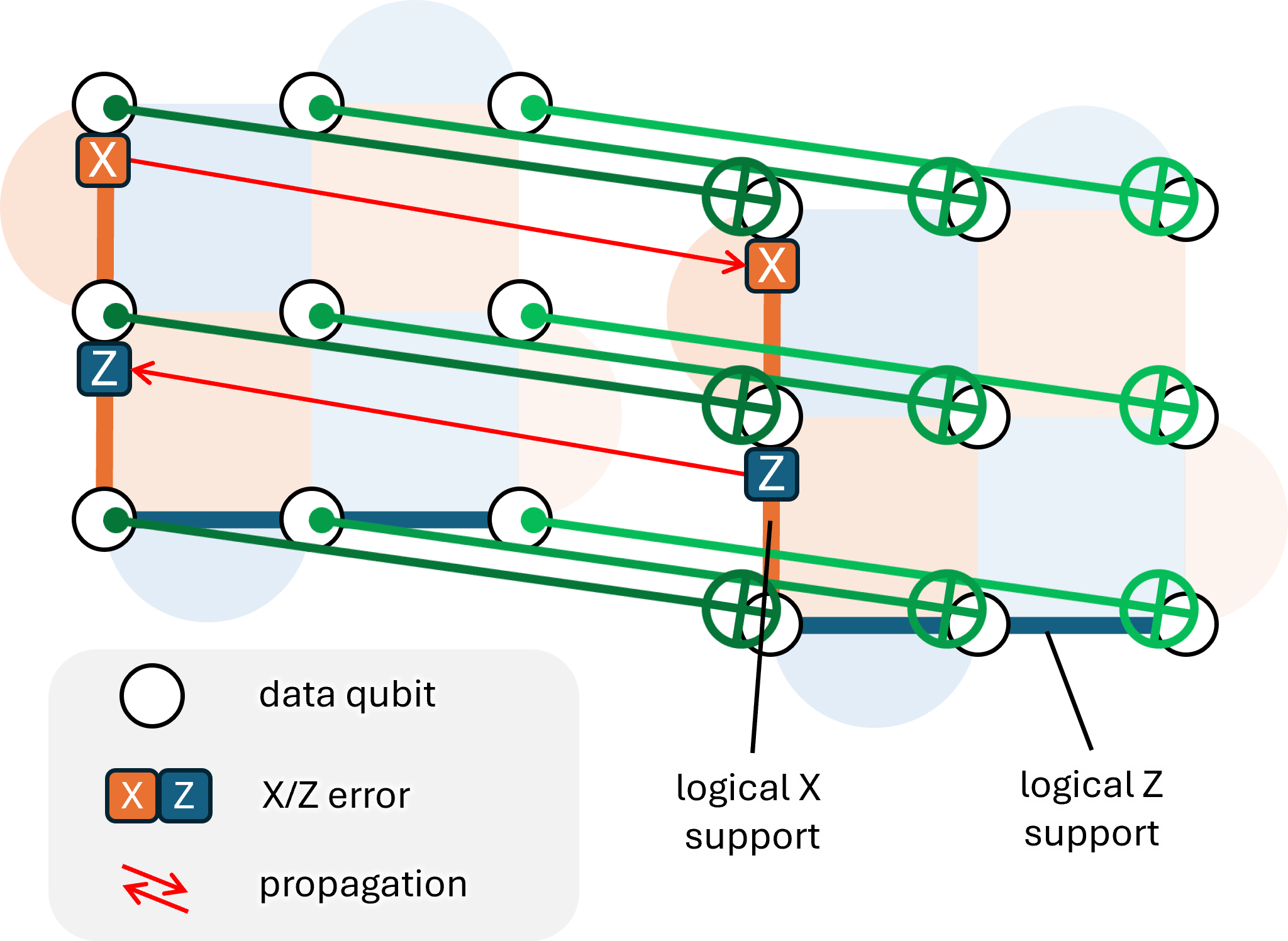}
    \caption{Schematic of a transversal CNOT between two distance-$3$ rotated surface-code patches. The deterministic propagation rules are shown for $X$-type errors, which propagate from control to target, and $Z$-type errors, which propagate from target to control.}
    \label{fig:tCNOT-gate-implementation}
\end{columnfigure}

The action of a single tCNOT is to conjugate the joint stabilizer group of two blocks, changing the relation between stabilizer measurements before and after the gate (see Appendix~\ref{app:stabilizer-transport}). The target block's $Z$ stabilizer values acquire contributions from the control, while the control's $X$ stabilizer values acquire contributions from the target. IRS does not use these transformed stabilizers, instead it forms detectors from consecutive measurements of the same local stabilizer as though the tCNOT was not present, as in a memory experiment. A detector event on one block can therefore arise either from a genuinely local fault, or due to propagation through a tCNOT. From the point of view of the memory-equivalent DEM, the latter is a source of contamination, introducing faults whose origin is not represented in the local error model. Figure~\ref{fig:error-prop} in Appendix~\ref{app:memory-stabilizers} illustrates this distinction.

Since a CNOT is self-inverse, this contribution can be deterministically undone by flipping the binary values of the corresponding detectors. We refer to this as the \emph{propagation correction}, resulting in sparser syndromes as the propagated components are stripped away, ideally leaving behind the genuinely local, memory-like part. However, in the presence of measurement faults, the measured stabilizer values are not sufficient to reliably identify data errors before the tCNOT. The local matching solution provided by MWPM is used to infer the configuration of faults immediately before a layer of tCNOTs, including the contribution of time-like matching edges that span the tCNOT gate boundary. The resulting correction is equivalent to transporting a locally inferred Pauli frame through the logical circuit~\cite{wan2024iterative}. This relieves the receiving block of an error burden not produced by its own local mechanisms and thus avoids assigning separate local explanations to the same propagated fault across blocks. We note that the propagation correction is retained separately from the local matching solution, and that removing its detector signature from the record does not remove it from the final solution provided by IRS.

\begin{figure}[htbp]
    \centering
    \includegraphics[width=\linewidth]{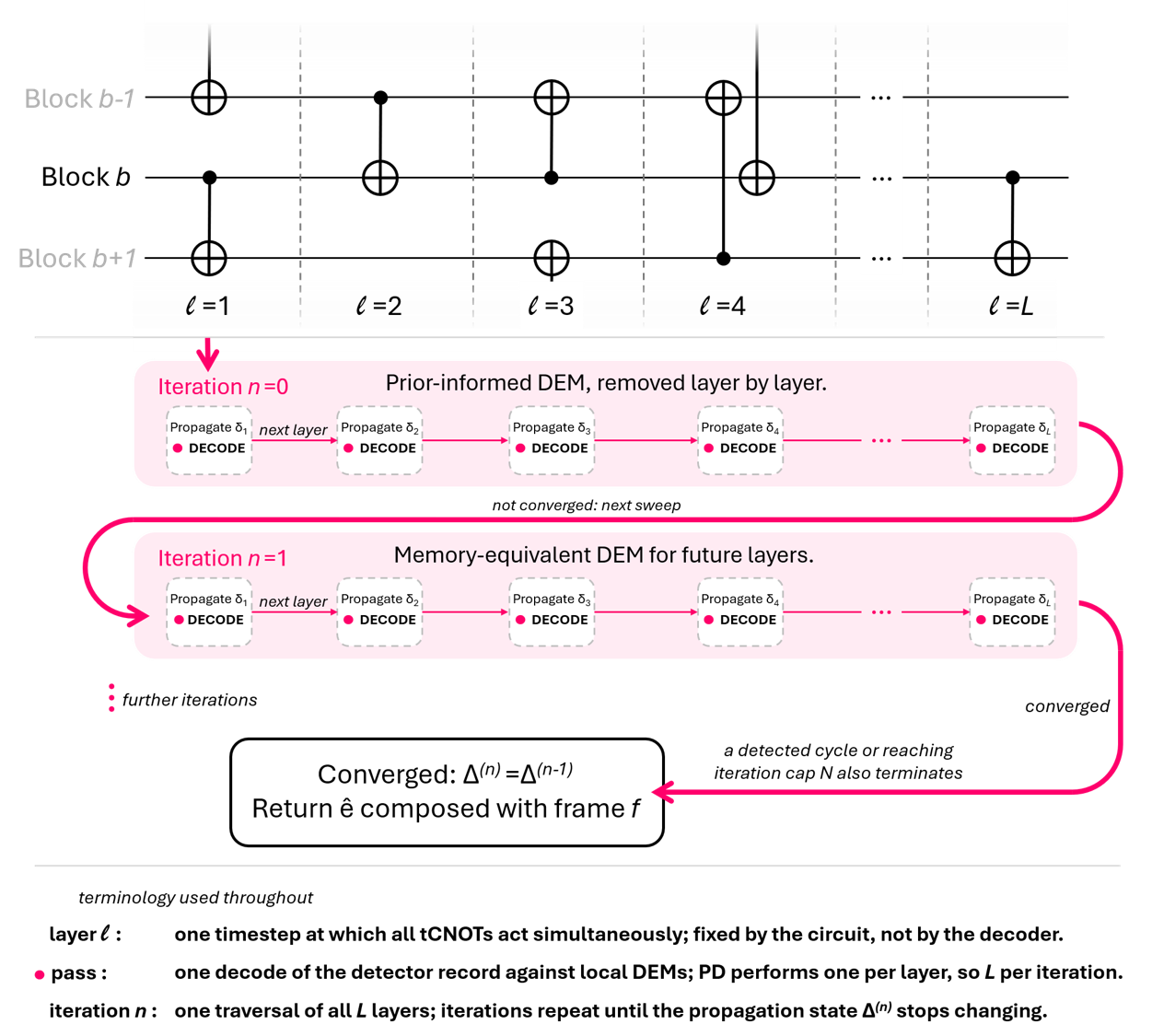}
    \caption{IRS layers, passes and iterations in the circuit being decoded, using the notation of Algorithm~\ref{alg:IRS}. Each layer $\ell$ contains simultaneous tCNOTs. Pink boxes represent one iteration: IRS visits the layers in circuit order, removes each layer's stored propagation correction $\boldsymbol{\delta}_\ell$, infers and applies an updated propagation correction, and re-decodes. Each iteration contains $L$ decoding passes; an initial pass precedes the first iteration. Iterations repeat until $\boldsymbol{\Delta}^{(n)}$ is unchanged between iterations, a cycle is detected, or the cap $N$ is reached. The prior-informed DEM used during the first iteration enhances the initial decoding estimates, and its contribution for each layer is removed once that layer's propagation correction has been applied. Subsequent iterations then use the bare memory-equivalent DEM. See Section \ref{sec:calibration} and Appendix \ref{app:prior-informed} for details.}
    \label{fig:pd-loops}
\end{figure}

The local decoding solution used to construct the propagation correction can itself be affected by the contamination it is intended to remove, so a single pass is not generally sufficient. For example, the first local solution may match a locally generated detector event to one introduced through propagation. Removing the inferred propagated contribution changes that matching problem, potentially changing both the preferred local explanation and the correction that should be propagated. To deal with this, IRS passes sequentially across logical-gate layers, so that each updated local solution informs the propagation correction applied to later layers. This is potentially repeated over multiple iterations over the full circuit, so that a propagation correction inferred at a later layer can revise an earlier local matching on the next iteration. This iterative flow is depicted in Figure~\ref{fig:pd-loops}, intended to disambiguate between four key terms: a \emph{block} is a logical-qubit patch; a \emph{layer} is a set of tCNOTs executed simultaneously; a \emph{pass} is one application of the local decoders to the detector records at a given timestep; and an \emph{iteration} visits all tCNOT layers in circuit order, sweeping through each layer to update the propagation correction and re-decode after each layer. For a circuit with $L$ tCNOT layers, there is one initial decoding pass followed by $L$ passes per iteration. The propagated Pauli frame is rebuilt during each iteration, while the stored detector corrections persist between iterations so that earlier updates can be undone and replaced.

For each tCNOT layer $\ell$, IRS stores a detector-toggle vector $\boldsymbol{\delta}_\ell$, initially zero. When the layer is revisited, its previous correction $\boldsymbol{\delta}_\ell^{\mathrm{old}}$ is removed and replaced by the newly inferred correction $\boldsymbol{\delta}_\ell^{\mathrm{new}}$, updating the working detector record $\mathbf{s}$ as
\begin{equation}
    \mathbf{s} \leftarrow \mathbf{s}
    \oplus \boldsymbol{\delta}_{\ell}^{\mathrm{old}}
    \oplus \boldsymbol{\delta}_{\ell}^{\mathrm{new}}.
\end{equation}
The new correction is stored and the detector record re-decoded, allowing propagation estimates to be revised rather than accumulated across iterations. The core procedure is as follows:
\begin{enumerate}
    \item \textbf{Initial pass.} Decode the measured detector record independently on each block using its local DEM. Initialise the stored propagation correction $\boldsymbol{\delta}_\ell$ for every tCNOT layer to zero.

    \item \textbf{Sweep through the gate layers.} This is one full iteration. Starting with an empty propagated Pauli frame, visit each tCNOT layer $\ell$ in circuit order, alternating propagation updates with local decoding passes:
    \begin{enumerate}
        \item Undo the propagation correction stored for this layer from the previous iteration (nothing on the first iteration).
        \item Use the local solutions from the preceding pass, together with the propagated frame accumulated from earlier layers in this iteration, to infer the Pauli errors present immediately before the tCNOTs. Propagate the relevant components through the gates, update the propagated frame, and generate a new detector correction $\boldsymbol{\delta}_\ell$.
        \item Apply and store the new correction, then perform a new decoding \emph{pass} across all affected blocks. The updated local solutions inform propagation at the next layer.
    \end{enumerate}
    Each iteration therefore contains $L$ decoding passes, one after each gate layer.

    \item \textbf{Repeat iterations.} Revisit all layers in circuit order, rebuilding the propagated Pauli frame each time. The stored detector corrections persist between iterations, allowing previous estimates to be replaced rather than accumulated. Stop when the ordered sequence of propagation corrections $\boldsymbol{\Delta}^{(n)}=(\boldsymbol{\delta}_1,\ldots,\boldsymbol{\delta}_L)$ is unchanged between successive iterations, a previous sequence recurs (indicating a cycle), or the maximum number of iterations is reached.

    \item \textbf{Final correction.} Combine the local error estimate from the final decoding pass with the propagated Pauli frame from the final iteration to obtain the complete error estimate and predict the logical observables.
\end{enumerate}

The explicit algorithm, including the optional prior-informed DEM described below, is provided in Appendix~\ref{app:algorithm}.

As IRS iterates, it explores a configuration space of local error estimates and their propagation corrections.
Before convergence, the local estimates and their implied corrections need not be consistent with the original global detector record.
However, at convergence, the final configuration will be mutually consistent and will explain the original detector record.
The Pauli frames of the final matching and the final propagation correction are then combined to obtain the decoding correction in order to predict the logical observables.

For a given detector record, IRS is deterministic: decoding the same initial detector record multiple times produces identical results, either arriving at the same fixed point in the same number of iterations or not converging.
It does not compare alternative globally consistent solutions to select the one with the highest global likelihood, as it lacks a single global objective function with which to make comparisons, and deterministically only generates a single solution rather than multiple ones, against which comparisons could otherwise be made.
Solving such global problems, such as \emph{loopy max-product} or \emph{belief-propagation}, efficiently is a longstanding optimisation problem~\cite{globerson2007fixing,sontag2009tree}, which IRS avoids altogether. 
As a result, convergence establishes self-consistency, guaranteeing that a \textit{coordinate-wise stationary point} has been found~\cite{yuan2024block}, but does not guarantee that the most likely global solution has been found.

IRS may fail to converge to a self-consistent solution, instead reaching an $n$-cycle rather than a fixed point; when this is detected, the decoder terminates and returns its latest logical prediction. 
Section~\ref{sec:iteration-dependence} quantifies the convergence and cycling behaviour of IRS and the impact of early termination  on logical performance.

All error-model details are given in Appendix~\ref{app:noise-models}, with details of circuit construction in Appendix \ref{app:circuit-construction}.

\subsection{Extensions}\label{sec:extensions}
The quality of the initial decoding estimates can be improved with an initial calibration step, using a \emph{prior-informed DEM}. 
Before any propagation correction has been computed and applied, the detector record contains the maximum amount of contamination from errors being propagated through the circuit. 
The memory-equivalent DEM does not contain mechanisms to explain this additional error burden, so the initial local matching is predisposed to favour less likely combinations of local faults to describe it. 
This, in turn, provides a misleading propagation correction to the next layer. 
The construction of a prior-informed DEM takes the existing memory-equivalent DEM and augments it with local, graph-like error mechanisms engineered to represent the contamination associated with tCNOT layers that have not yet been processed. 
These contributions are removed from the DEM layer by layer as the first iteration proceeds, and the corresponding propagation corrections are applied explicitly. 
By the end of the first full iteration, the model has returned to the bare memory-equivalent DEM, which is used throughout all subsequent iterations. 
The propagation-and-redecoding rule and its potential fixed points are unchanged, but the altered initial cost landscape of the local decoders can place IRS in the basin of attraction of a different fixed point. 
We find this increases the probability of reaching a fixed point with the correct logical prediction. 
The underlying IRS procedure can be run without this calibration step; the construction of the \emph{prior}, here referring to a modified error model which accounts for propagated errors, and its effect on decoding performance are studied separately in Section~\ref{sec:calibration}.

Retaining the memory-like detectors used by IRS has both advantages and computational costs. Measurement errors retain their ordinary local, time-like detector signatures, including across tCNOT boundaries, and the local decoder solution preserves the space-time information needed to infer the pre-gate error configuration. As the local decoding solution changes, IRS can revise which inferred errors were present before the gate and update their propagated consequences. 

For real-time operation, IRS would need to be implemented in a windowed form where decoding is performed over a bounded span of the recent syndrome history rather than over the full circuit record. The detector volume associated with the propagated events would be bounded by the window size, rather than increasing with total circuit depth, with propagated information that reaches a temporal window boundary being retained in a compact boundary state, for example through the accumulated propagated frame or equivalent edge information, and transferred consistently into the next window. 
This suggests that the memory-like detector representation is compatible with cost-bounded real-time decoding, even in deep circuits, provided that the boundary state is handled correctly. A windowed streaming extension of an earlier form of IRS was already proposed in Ref.~\cite{wan2024iterative}, including a concrete algorithm and worked demonstration of its operation on streaming syndrome data. More extensive implementation, numerical validation, and comparison of the resulting logical-performance and latency trade-offs are left to future work.

In contrast to the memory-like detectors used by IRS, pre-gate and post-gate detector frames incorporate stabilizer transport directly into their definitions, making propagated data-error signatures more local and producing a sparser detector record. The trade-off is that some detector definitions span measurements across logical blocks, shifting part of the non-local structure into measurement-error signatures~\cite{serra2026decoding}. These gate aware detector frames are not directly compatible with IRS, since they build stabilizer transport into fixed detector definitions, whereas IRS keeps propagation separate so that its inferred contributions can be iteratively refined as the local decoding solutions improve.

We note that the repeated local decoding calls and message passing inherent to IRS add further overhead, with the decode calls growing linearly per-layer. However, each local decoding problem nevertheless retains the size and structure of the corresponding memory experiment of the same duration, rather than becoming a joint cross-block decoding problem.

\section{Robust Handling of Indeterminate Initial Stabilizers}\label{sec:initial-stabilizers}
\begin{widefigure}[ht]
    \centering
    \includegraphics[width=.9\linewidth]{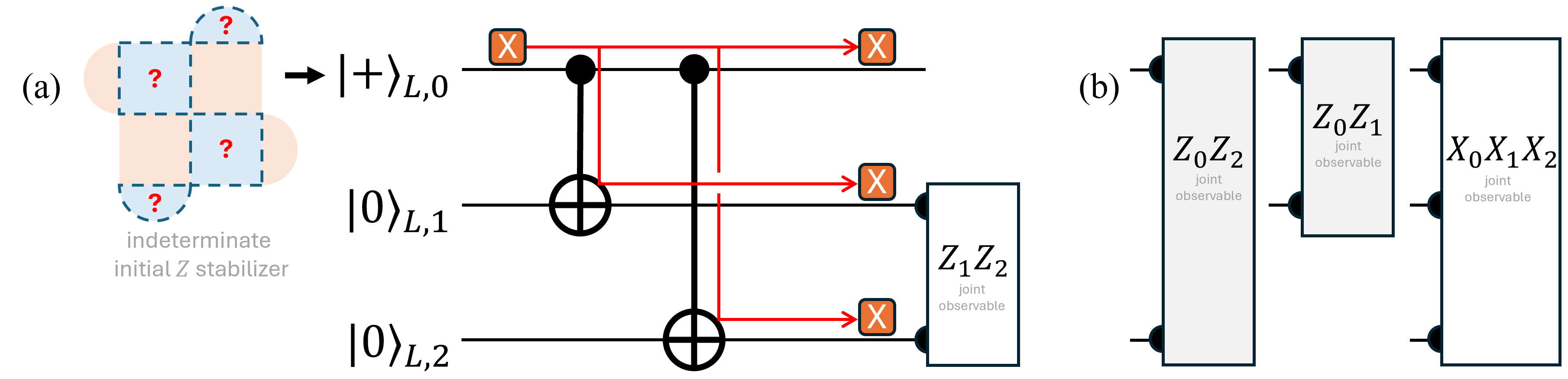}
    \caption{(a) Logical circuit for preparing a 3 logical qubit GHZ state. The $0^\textnormal{th}$ qubit is initialised in the logical plus state, for which the initial $Z$-type stabilizer values are indeterminate, indicated by red question marks on the blue stabilizers; this is the standard gauge freedom when initialising in the $Z(X)$-basis and measuring the $X(Z)$-stabilizers. This gauge propagates across the tCNOTs onto other blocks, represented by the $X$ enclosed in an orange box, whose contributions cancel in the logical $Z_1Z_2$ observable. The resulting performance is shown in Figure~\ref{fig:quera_ghz_graph}.
    (b) Other deterministic logical measurements on the GHZ-state. The contributions from the initially indeterminate $Z$-stabilizer outcomes do not cancel for the greyed-out measurements, so must be handled directly within IRS.}
    \label{fig:ghz-circuit}
\end{widefigure}

Ordinary memory decoding does not define detectors using the first measurement of an initially indeterminate stabilizer. That initial measurement instead forms part of state preparation, serving to project the state into the codespace and establish the reference measurements against which detectors are defined in subsequent rounds. For example, initialising  data qubits in the surface code to $\ket{0}$ states fixes every $Z$ stabilizer and the logical $Z_L$ to be $+1$, but leaves the $X$ stabilizers indeterminate. This product state is not yet in the codespace until the X stabilizers are also measured; the first stabilizer measurements project into a definite stabilizer (parity) eigenspace with uniformly random $X$ stabilizer outcomes. These outcomes establish the initial syndrome frame, which we refer to as the \emph{preparation gauge}. Subsequent memory rounds repeat the same values in the absence of errors. Exchanging $X$ and $Z$ gives the corresponding construction for $\ket{+}_L$.

This convention is sufficient for an isolated memory experiment, but is not compatible with IRS. If no detector is defined for the first measurement of an initially indeterminate stabilizer, its random value is absent from that block's detector record. Those stabilizer values can subsequently be transported through tCNOTs onto another block without a valid determinate cross-block detector being defined. If memory-like detectors are used, this transported stabilizer generates detector events on receiving block, appearing as propagated contamination without a corresponding source mechanism available to the decoder. 

The GHZ circuit depicted in Figure~\ref{fig:ghz-circuit} illustrates why this preparation randomness is problematic for some decoding approaches. The random measurement outcomes of an initially indeterminate stabilizer set the preparation gauge, which is propagated through the tCNOT entangling circuit, stabilises the ideal GHZ state and leaves its deterministic joint observables unchanged. For IRS, without an explicit representation of the preparation gauge, the resulting high-weight detector patterns propagated onto later blocks must be independently explained by local decoders. In the presence of genuine faults, those explanations may not remain mutually consistent, resulting in logical failure by the decoder. The same applies to the Bell-state preparation circuit, for which, even under noiseless execution, no choice of valid logical observable avoids this failure mode without the appropriate treatment.

The general solution is to include the initially indeterminate stabilizer outcomes in the memory-like detector record and allow IRS to infer their propagated contribution. Since these outcomes are random even in the absence of noise, any detectors defined from them alone are \emph{gauge detectors}, and Stim rejects the corresponding DEM by default. We generate the memory-equivalent DEMs with \texttt{allow\_gauge\_detectors=True}, which represents this preparation randomness through probability-$1/2$ gauge mechanisms~\cite{stim_docs,gidney2021stim} with zero matching weight, allowing local decoders to explain the random measurements without contributing to the weight of the matching. They remain in the memory-equivalent model throughout the IRS iterations. IRS then maps the matched gauge edges, including paths to the code boundaries, to Pauli chains on the data qubits which supply the spatial error configurations needed to predict the propagated detector contribution, and propagates them through the circuit. These Pauli chains are degenerate; several chains can explain the same initial syndrome, with the requirement that the chosen chain and inferred physical faults together reproduce the local detector record. Once transported, the inferred gauge contribution supplies the updates used to remove its propagated contamination from the receiving blocks.

Figure~\ref{fig:quera_ghz_graph} compares logical GHZ state preparation using IRS with the published results of Ref.~\cite{cain2025fast}, using its $3$-block GHZ logical circuit and physical error model. The black trend is taken from Figure~S7 of that work; the \emph{independent patch-wise error transfer} decoder is their implementation of a PAL decoder associated with several earlier decoding approaches, including Ref.~\cite{wan2024iterative}. In contrast to the published failure of their PAL implementation, IRS continues to suppress logical errors with increasing distance on the same circuit, demonstrating that this failure is not an intrinsic limitation of PAL decoding in general.

\begin{columnfigure}[ht]
    \centering
    \includegraphics[width=\linewidth]{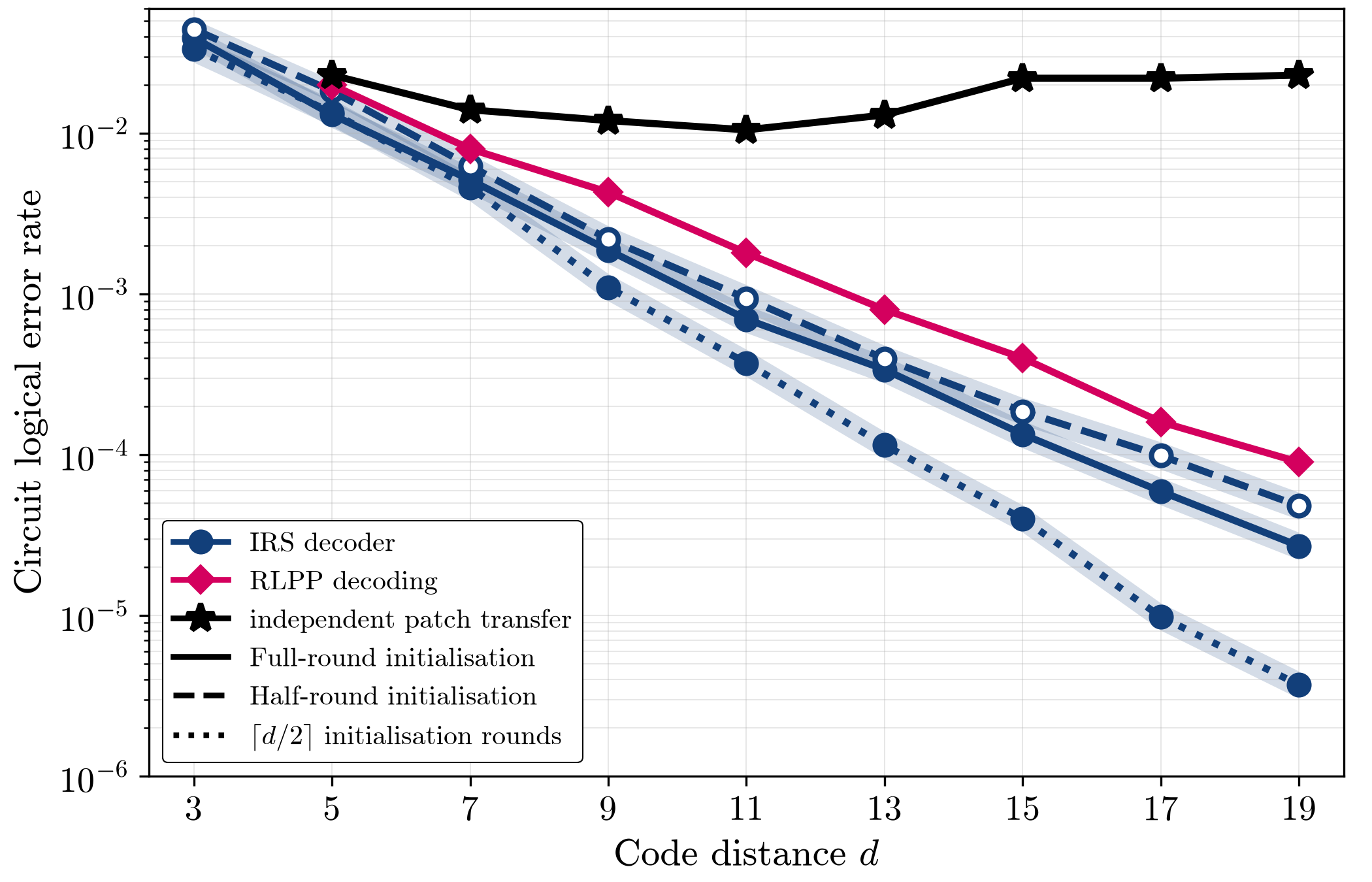}
    \caption{$3$-block GHZ benchmark using the logical circuit and physical error model of Ref.~\cite{cain2025fast}. Circuit logical error rate is shown versus code distance $d$ for IRS with correlated MWPM (blue circles), the published reliable logical Pauli-products (RLPP) decoder (magenta diamonds), and the independent patch-transfer reference (black stars) from Figure~S7 of Ref.~\cite{cain2025fast}. Solid lines with filled markers, and dashed lines with empty markers, use full-full and half-full initialisation, respectively (see main text). The dotted blue curve uses $\lceil d/2\rceil$ full syndrome-extraction rounds before the first tCNOT. Shaded bands show 95\% confidence Wilson intervals. Error-model details are given in Appendix~\ref{app:noise-models}. Additional LOM comparisons are presented in Figure~\ref{fig:ghz-additional} of Appendix~\ref{app:ghz-decoding}.}
    \label{fig:quera_ghz_graph}
\end{columnfigure}

We also compare with reliable logical Pauli product (RLPP) decoding~\cite{cain2025fast}, which shares the same basic principle of logical observable matching (LOM)~\cite{serra2026decoding}: back-propagating a reliable logical observable through the circuit and restricting matching to the associated stabilizer information. We tested this correspondence using the Delft open-source LOM implementation~\cite{SerraPeralta_LOM} on the reference GHZ circuit, shown in Appendix \ref{app:ghz-decoding}, with the noise model and initialisation schedule of Ref.~\cite{cain2025fast}~\footnote{The reference circuit and error model were provided directly by the authors of Ref.~\cite{cain2025fast} in private communication.}.
The reference circuit from Ref.~\cite{cain2025fast} uses a restricted \emph{half-round} initialisation schedule, suitable for RLPP decoding. The first round measures the initially indeterminate stabilizer family, for example the $Z$ stabilizers of the plus-state logical block, and not the other basis, followed by a second full SE round measuring both stabilizer bases. Using this half-round initialisation, IRS without correlated MWPM achieves logical error rates comparable to the reliable logical Pauli-products (RLPP) decoder of Ref.~\cite{cain2025fast}. Enabling correlated MWPM~\cite{higgott2025sparse} for IRS improves decoder quality further: at $d=19$, IRS gives a logical error rate of approximately $4.3\times10^{-5}$, compared with $\sim 9\times10^{-5}$ for the RLPP decoder.
In the presence of noise, the omitted measurements not present in the half-round still provide useful information for IRS about the preparation gauge. A \emph{full-full} SE schedule where both $X$ and $Z$ stabilizers are measured in both rounds results in a further reduction in logical error rates for both IRS and the open-source LOM decoder~\cite{serra2026decoding}, which follows the same underlying principle as RLPP decoding. 
The benefit is therefore not exclusive to IRS: both approaches can exploit the early syndrome information omitted by half-round initialisation. 

Increasing the number of full syndrome-extraction rounds before the first tCNOT to $\lceil d/2\rceil$ full SE rounds drives a further reduction in logical error rates for IRS, with the correlated IRS curve in Figure~\ref{fig:quera_ghz_graph} exhibiting steeper suppression as a function of distance than either fixed-initialisation curve. At $d=19$, ten full initialisation rounds give a logical error rate of approximately $3.6\times10^{-6}$, a factor of $25$ below the best available result for the RLPP decoder. 
We observe the same behaviour for Bell state preparation circuits in Appendix~\ref{app:ghz-decoding}. With $\lceil d/2 \rceil$ initialisation rounds, a circuit initialised in $\ket{+}_L\ket{0}_L$ followed by a tCNOT reaches logical error rates indistinguishable from those of the same circuit with $\ket0_L\ket0_L$ initialisation, for which there is no relevant preparation gauge for IRS to track. This supports the conclusion that appropriate state preparation can suppress the decoding penalty induced by the preparation gauge in circuits with relevant indeterminate initial stabilizers.

We understand the different initialisation requirements and resulting performance differences between methods in the following way.  
For RLPP decoding, back-propagating the $Z_L^{(1)}Z_L^{(2)}$ observable gives $Z_L^{(0)}Z_L^{(1)}Z_L^{(2)}$ between the two tCNOTs and $Z_L^{(1)}Z_L^{(2)}$ before the first. The plus-state block (logical qubit $0$) has no support in the back-propagated observable during initialisation, so for both RLPP and LOM the $Z$-stabilizers of that block are never used to generate detectors involving only that block~\cite{cain2025fast,serra2026decoding}. These stabilizers can still enter cross-block detector definitions where they occur directly adjacent to a tCNOT, but detectors involving only the plus-state block are entirely absent. In contrast, IRS retains these detectors to help distinguish between time-like (measurement) and space-like (data) matchings whose error configurations would propagate differently through the tCNOTs, using them when inferring which configuration to propagate.

\section{Iteration Dependence and Cycling}\label{sec:iteration-dependence}
\begin{widefigure}[ht]
    \centering
    \includegraphics[width=\linewidth]{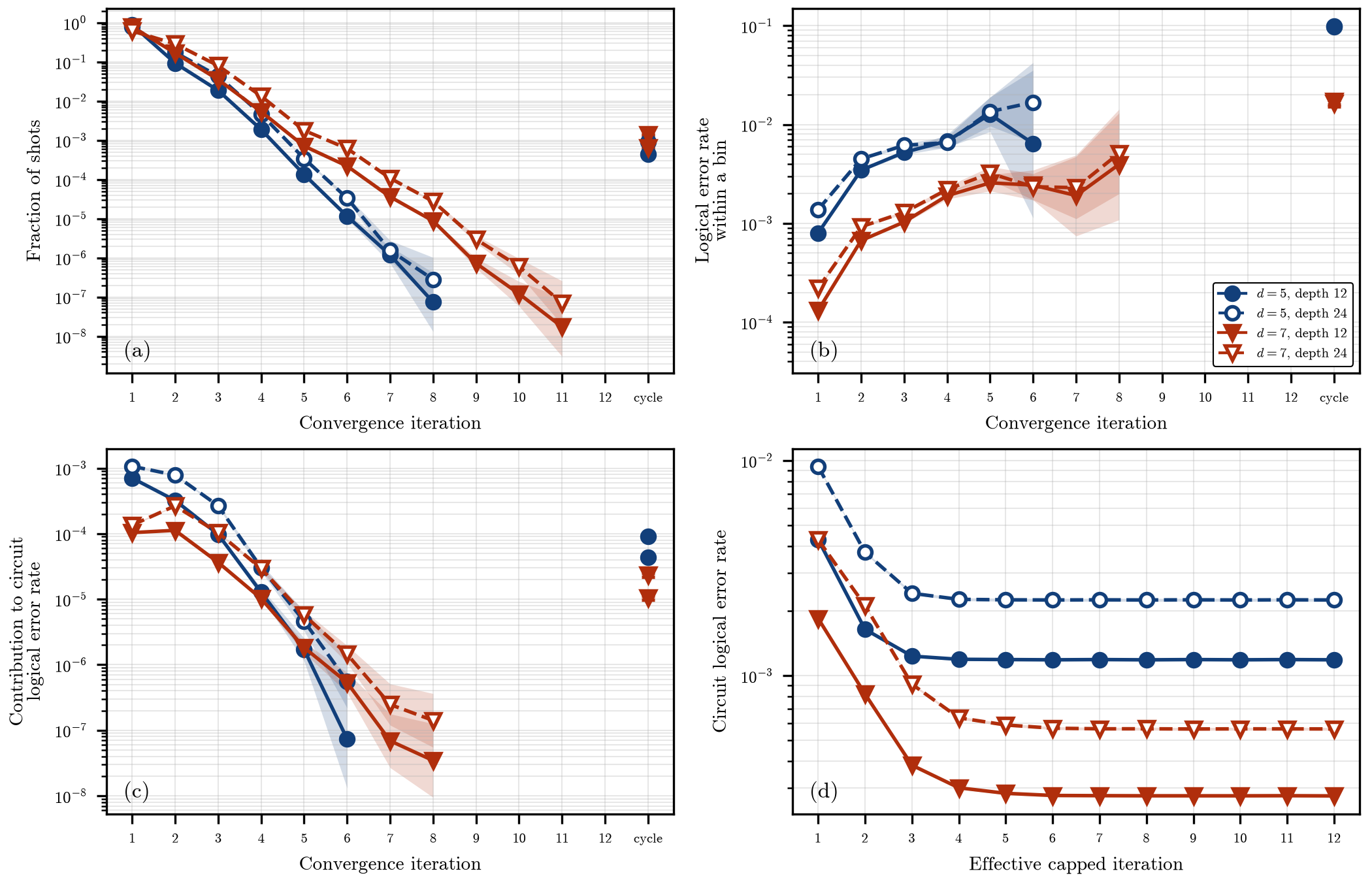}
    \caption{Iteration dependence and cycling of IRS for alternating-tCNOT circuits at depths $12$ and $24$. Here, depth refers to the logical depth, that is, the depth of tCNOT gates. It does not include the one SE round between each tCNOT. Logical blocks are initialised in $\ket{0}_L$, with each block measured individually in the $Z$-basis. Blue and red trends show $d=5$ and $d=7$, respectively; filled circles with solid lines indicate depth $12$, and open circle with dashed lines indicate depth $24$. Each iteration is a complete circuit sweep, including the local redecoding and propagation updates at successive layers. Shots are grouped by the iteration in which they converged, with nonconvergent shots assigned to the `cycle' bin. (a) Fraction of all shots in each bin. (b) The logical error rate obtained for shots which converged after a specific number of iterations, i.e., the logical error rate within a bin. (c) Contribution of each bin to the overall logical error rate, calculated as the number of failures in that bin divided by the total number of shots. (d) Circuit logical error rate versus the effective maximum iteration count. Missing points in sparsely populated bins in panels (a)-(c) reflect finite sampling rather than a sharp convergence cutoff. Shaded bands show 95\% confidence Wilson intervals.}
    \label{fig:iteration_dependence}
\end{widefigure}

For the majority of circuits, the IRS solution will converge with an error rate close to that of the corresponding memory decoding problem.
However, in some cases the solution will instead be a cycle, oscillating between different matchings.
We study the alternating tCNOT case because it represents a uniquely demanding class of circuits in which a sequence of propagation corrections can repeatedly return a Pauli error back to the block from which it originated. 
For example, an $X$-error estimate on block $A$ is propagated to $B$ by a $CX_{A\rightarrow B}$, changing the local decoding problem on $B$. 
A subsequent, closely spaced $CX_{B\rightarrow A}$ (i.e., $\leq d/2$ SE rounds between them) then propagates the $X$ error back to $A$, and the cycle repeats. 
In general, any circuit containing error feedback between logical blocks will require more than one iteration to converge to a stable decoding solution, while some of them will enter a cycle.
The alternating pattern studied here extends this feedback across many loops, providing a deliberately adversarial test of IRS. It is not intended to represent typical transversal Clifford circuits.

In Figure~\ref{fig:iteration_dependence}, we examine the convergence properties of IRS for chains of alternating-tCNOTs of depths $12$ and $24$, for code distances $d=5$ and $d=7$. 
Increasing the number of SE rounds between alternating tCNOTs helps determine whether edges should be matched before, across, or after the gates, resulting in different propagation corrections in each case. 
With sufficiently large $O(d)$ separation, the alternating circuits studied can be decoded in only a single iteration, with no cycling. 
This comes at the cost of a distance-dependent time overhead between logical gates, forfeiting the constant-round gate separation that makes fast transversal logic attractive, so we focus in Figure~\ref{fig:iteration_dependence} how well IRS performs with only a single SE round between tCNOTs.
We group shots into \textit{bins}: grouped by the number of complete iterations required for the propagation state $\bm{\Delta}^{(n)}$ to stop changing. Numbered convergence bins appearing on the horizontal axes of Figures~\ref{fig:iteration_dependence}(a)-(c) contain only those shots which converged after that many iterations, while the separate \emph{cycle} bin contains shots which entered a cycle (an earlier propagation state recurs) without convergence. 

Panel~(a) shows that $\gtrsim90\%$ of shots converge after only one or two iterations, and that these shots have the smallest average contribution to the logical error rate, as can be seen in panel~(b). 
Among shots that converge, the population decreases exponentially with the number of iterations required, falling by roughly an order of magnitude per additional iteration. 
The mean number of iterations for convergence is therefore dominated by the first few bins, even though a small number of shots require additional iterations. 
The cycling population behaves differently: it accounts for roughly $0.1\%$ of the total number of shots, a far larger fraction than would be predicted by extrapolating the trend asymptotically towards ever more iterations.
This follows from the discrete nature of the deterministic update rule: a complete iteration is a fixed deterministic update of the propagation configuration, assuming deterministic local decoding.
For a fixed circuit, this configuration can only take a finite number of states, so every circuit will always reach either a fixed point or a cycle during decoding. 
Since the update rule is deterministic, repetition of an earlier state guarantees that the decoder has entered a cycle, so provides a direct cycle-detection condition.

Panel~(c) reports each bin's contribution to the overall logical error rate, defined as the number of logical failures in that bin divided by the total number of shots. 
The sum of every bin, including the cycle bin, therefore recovers the overall logical error rate. 
The population of a bin (panel~(a)) falls with iteration count roughly five times as fast as its logical error rate per bin (panel~(b)) rises (on a log scale), so each successive bin contributes close to an order of magnitude less to the overall error rate than the last: shots that take longer to converge are exponentially suppressed in their effect on the logical error rate. 
Shots which enter a cycle contribute two to three orders of magnitude more than those in the highest-numbered convergence bins, but still roughly an order of magnitude less than those converging in one or two iterations. 
The dominant contribution therefore comes from those shots that converge in only one or two iterations, and not from those which enter a cycle.

In light of this, panel~(d) imposes an explicit cap on the number of iterations, returning the current decoding solution when the cap is reached. 
The logical error rate falls rapidly over the first few iterations before plateauing after 3-4 iterations for $d=5$ and 5-6 iterations for $d=7$, with the required number of iterations growing with code distance while being largely independent of circuit depth. 
This distance dependence follows from the fact that larger distance blocks require a longer chain of faults to cause a logical error, which is itself more likely to span a larger number of layers.
The decoder then needs more iterations to correctly disambiguate between genuine chains of faults and those which include propagated contamination, as in panel~(a) where a greater fraction of shots require more iterations to converge at distance 7 than 5.

Raising the cap beyond the emergence of the plateau yields negligible further improvement. 
The gain from one additional iteration is bounded by the contribution of the corresponding bin in panel~(c), being exponentially suppressed and falling by close to an order of magnitude per iteration, so the available improvement is exhausted within the first few iterations.
Since the cap terminates every shot not already converged, including those in a cycle, the flatness of the plateau indicates that this forced termination does not result in a measurable degradation in performance. 
Explicit cycle detection is therefore not required for logical accuracy, but rather in reducing decoder runtime by detecting shots which have already entered a cycle and can be terminated before reaching the iteration cap.

We conclude that it is often sufficient to terminate IRS iterations early, either by an iteration cap or with cycle-detection, while largely retaining its decoding performance. 
This avoids wasting limited decoding time on a negligible improvement in logical performance.
We again note that long chains of alternating tCNOTs represent an especially adversarial class of circuits, being the worst case among all constructions we have tested. This suggests that more representative circuits will require only a very small iteration cap relative to the code distance, without needing the large syndrome-extraction separation between tCNOTs that would make a single iteration sufficient.

\section{Prior-Informed DEM}\label{sec:calibration}
Consider a circuit containing tCNOTs but being decoded using the memory-equivalent DEM in which all hypergraph-like edges have been removed. 
The detector record on each logical block contains two types of faults: (i) genuinely local faults and faults introduced directly by the noisy physical CNOTs, as well as (ii) contamination caused by propagation from other logical blocks. 
The first category of faults is well-described by the mechanisms contained within the DEM, whereas the second category is not, and the decoder is forced to find a matching using the local mechanisms available to it. 
This results in high-weight matchings which correspond to combinations of errors with a very low probability, effectively biasing MWPM towards unreasonable error configurations.

This describes the first iteration of IRS: based on these poor initial matchings, IRS generates a propagation correction, which infers which detectors fired due to propagation and which corresponded to genuinely local faults, and then updates the detector record accordingly ready for the next iteration. 
This is essentially a bootstrap problem: the propagation correction improves the local matching, but its own quality depends on the accuracy of the matching which produced it, and the initial matching had no propagation correction. 
We can improve this situation by modifying the error model used during the first iteration, leaving the decoder, the detectors, and the matching machinery unchanged.
This modified error model is not intended to reproduce the correlated detector structure of the transversal circuit (which is what the propagation correction itself does), but instead provides the local decoder with a re-weighted approximate explanation for those detector events which are caused by propagation.
We refer to the resulting DEM as the \emph{prior-informed DEM}, since it contains prior knowledge of the expected contamination that is likely to occur due to propagation across the CNOTs.

\begin{widefigure}[t]
    \centering
    \includegraphics[width=\linewidth]{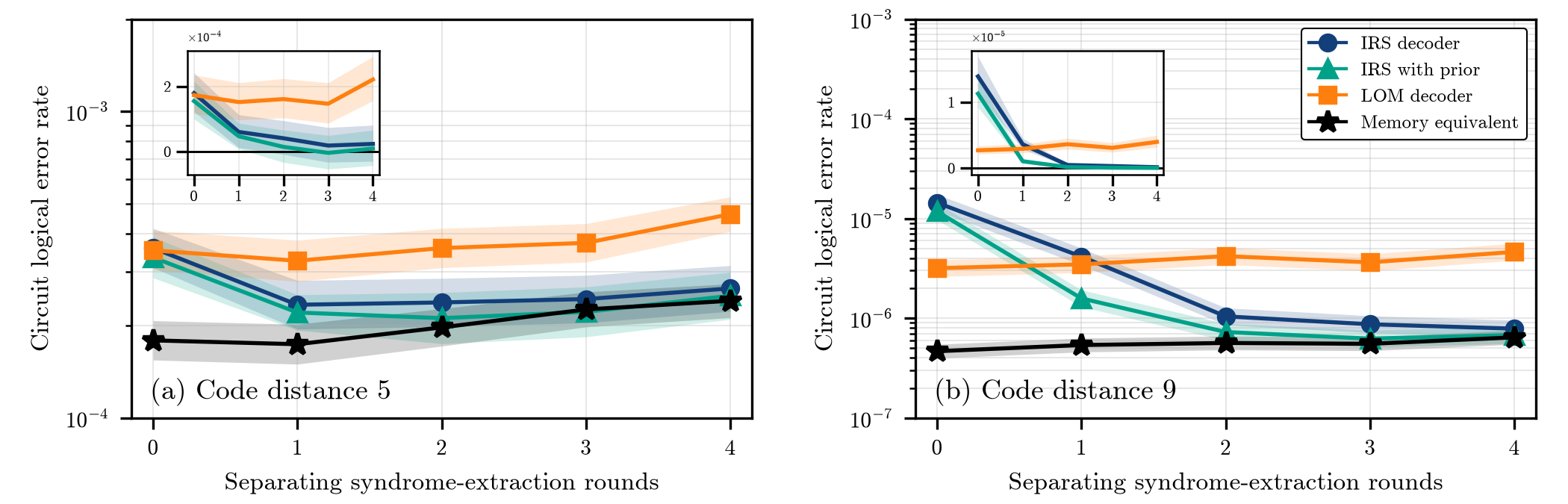}
    \caption{Logical error rate for an alternating-tCNOT pair, varying the number of separating SE rounds at code distances (a) $d=5$ and (b) $d=9$. Logical blocks are initialised in $\ket{0}_L$, with each block measured individually in the $Z$-basis. Blue trends show IRS, teal trends show IRS with a prior-informed initial pass, orange trends show LOM, and black trends show the memory experiment. Insets show logical error rates minus the memory baseline at the same buffer-round count. Shaded bands indicate the 95\% confidence Wilson interval.}
    \label{fig:calibration}
\end{widefigure}

The expected contamination is obtained by propagating the physical error model forward in time (using the appropriate commutation relations) through the layers of the logical circuit and updating the probability of observing an error ($X$ or $Z$) for each data qubit after each gate. 
After a CNOT, the probability of an $X$ error on the target data qubit acquires the probability imported from the control data qubit, and the probability of a $Z$ error on the control data qubit acquires the probability imported from the target.
Probabilities are combined by parity, with two identical Pauli components cancelling to the identity, e.g. $XX=I$ (see Appendix \ref{app:prior-informed} for details).
Repeating this through the remaining layers gives the expected local contamination at every layer of the circuit which is appended to the memory-equivalent DEM to be used when decoding each layer; this is the prior-informed DEM, which still only contains local error mechanisms, but has been re-weighted according to how faults are expected to be detected, rather than where they are expected to occur.
Writing $p_{\mathrm{acc}}$ for the error the data qubit has already accumulated before the gate, $p_{\mathrm{prop}}$ for the contribution propagated across the gate, and $p_{\mathrm{dir}}$ for the noise associated with the physical tCNOT, the added mechanism $q$ is determined by requiring that it reproduces the contamination expected immediately after the gate,
\begin{align}\label{eq:PAR}
    &\mathrm{PAR}(p_{\mathrm{acc}}, q) = \mathrm{PAR}(p_{\mathrm{acc}}, p_{\mathrm{prop}}, p_{\mathrm{dir}}) \nonumber \\
    &\qquad \Longrightarrow \quad
    q = \mathrm{PAR}(p_{\mathrm{prop}}, p_{\mathrm{dir}}).
\end{align}
This follows from the fact that $\mathrm{PAR}$ (parity) composes associatively, so $p_{\mathrm{acc}}$ cancels identically and the added mechanism depends only on what the gate itself introduces.

Having obtained the prior-informed DEM, IRS begins its first iteration through the circuit. 
At each layer, a matching is inferred in the normal way, the propagation correction for that layer is applied to the detector record, and the prior contribution associated with that layer is then removed from the DEM.
This removal avoids double counting: once the propagation across a layer has been explicitly inferred and corrected for, the same contamination would otherwise be represented both in the corrected detector record and in the error model used to explain it.
By the end of the first iteration the decoder has returned to the unaugmented memory-equivalent DEM, which is then used in every subsequent iteration.
The prior therefore changes only the bootstrap into the iteration, and leaves the propagation and redecoding rules themselves unchanged.

Figure \ref{fig:calibration} isolates the impact of the prior-informed DEM on IRS as a function of the number of SE rounds between tCNOTs.
For code distances $d = 5$ and $d = 9$, we vary the number of SE rounds separating consecutive tCNOTs and report the resulting logical error rate, alongside the memory-equivalent reference obtained by decoding the same blocks with the tCNOTs removed.
The reference serves as the natural target, since it is the error rate the circuit would incur absent the effects of propagation and the noisy CNOTs themselves. 
As expected, the logical error rate in the reference case itself increases with the number of separating rounds, as each additional round of syndrome extraction contributes additional measurement error.
IRS with and without the aid of the prior-informed DEM during the first iteration improve rapidly with separating SE rounds, whereas the LOM decoder remains largely insensitive to the additional information provided by more SE rounds 
At $d=9$ and zero separating rounds LOM performs better than IRS, since IRS both with and without the prior-informed DEM lacks the temporal information needed to place inferred errors on either side of the gates. 
This ordering reverses once that information becomes available: IRS with the prior-informed DEM outperforms LOM at one separating round and reaches the memory-equivalent reference at two, whereas standard IRS requires two rounds to outperform LOM and takes longer to reach the memory-equivalent reference. 
Including the prior-informed DEM is more impactful when the number of separating SE rounds is small, where the prior is the main source of information regarding which detector events are due to propagation. 
As more rounds are inserted between the gates, the extra syndrome information supplies that information instead, and IRS with and without the prior-informed DEM converge to the same memory-equivalent limit. 
This benefit also grows with distance: at $d=5$ the variants lie within error at every separation, whereas at $d=9$ the calibrated decoder is lower by a factor of nearly three at one separating round.

\section{Application to the S gate}\label{sec:S-gate}
The tCNOT is not the only Clifford gate that the surface code admits at constant depth, though other gates are harder to obtain. 
No surface code admits a strictly transversal $S$, but the unrotated code, which has roughly twice the physical qubits per unit distance, does admit a fold-transversal one \cite{moussa2016transversal,breuckmann2024fold}.
Two recent constructions obtain $S$ on the rotated surface code instead. 
Ref.~\cite{chen2026transversal} embeds a fold-transversal $S$ gate within a single syndrome-extraction round, applying it to the mid-cycle state when the data and ancilla qubits of a rotated patch jointly form an unrotated surface code \cite{mcewen2023relaxing}. 
Ref.~\cite{holmes2026quantum} instead performs a logical $S$ using data qubits only.
It uses a depth-one layer of $d$ $S$ gates, distributed mostly on qubits along the central diagonal of the block, with a single layer of $CZ$ gates on qubit pairs described below.

Both constructions require decoding support beyond standard memory-stabilizer decoding. 
We use the construction of Ref.~\cite{holmes2026quantum}, depicted in Figure \ref{fig:S-gate-implementation}.
Given qubit $(0,0)$ is in the top left corner, it applies $d-1$ phase gates on the diagonal qubits $(r,r)$ for $0 \leq r \leq d-2$, one phase gate at $(d-2,d-1)$, and $\binom{d}{2}$ $CZ$ gates pairing each qubit on the diagonal with its immediate neighbour to the right and each below-diagonal qubit with its reflection across the diagonal shifted one column to the right.
The controlled-$Z$ pairs are disjoint, so the layer is $2$-local and of depth one, and it occupies the same position between syndrome-extraction rounds as a tCNOT layer.
We note that Ref.~\cite{holmes2026quantum} does not address decoding this $S$ gate construction.
It establishes that the implementation is distance-preserving, and bounds the circuit-level distance of the depth-one gadget between $\lceil d/2 \rceil$ and $d-1$, making it single-fault-tolerant only for $d \geq 5$.

\begin{columnfigure}[ht]
    \centering
    \includegraphics[width=\linewidth]{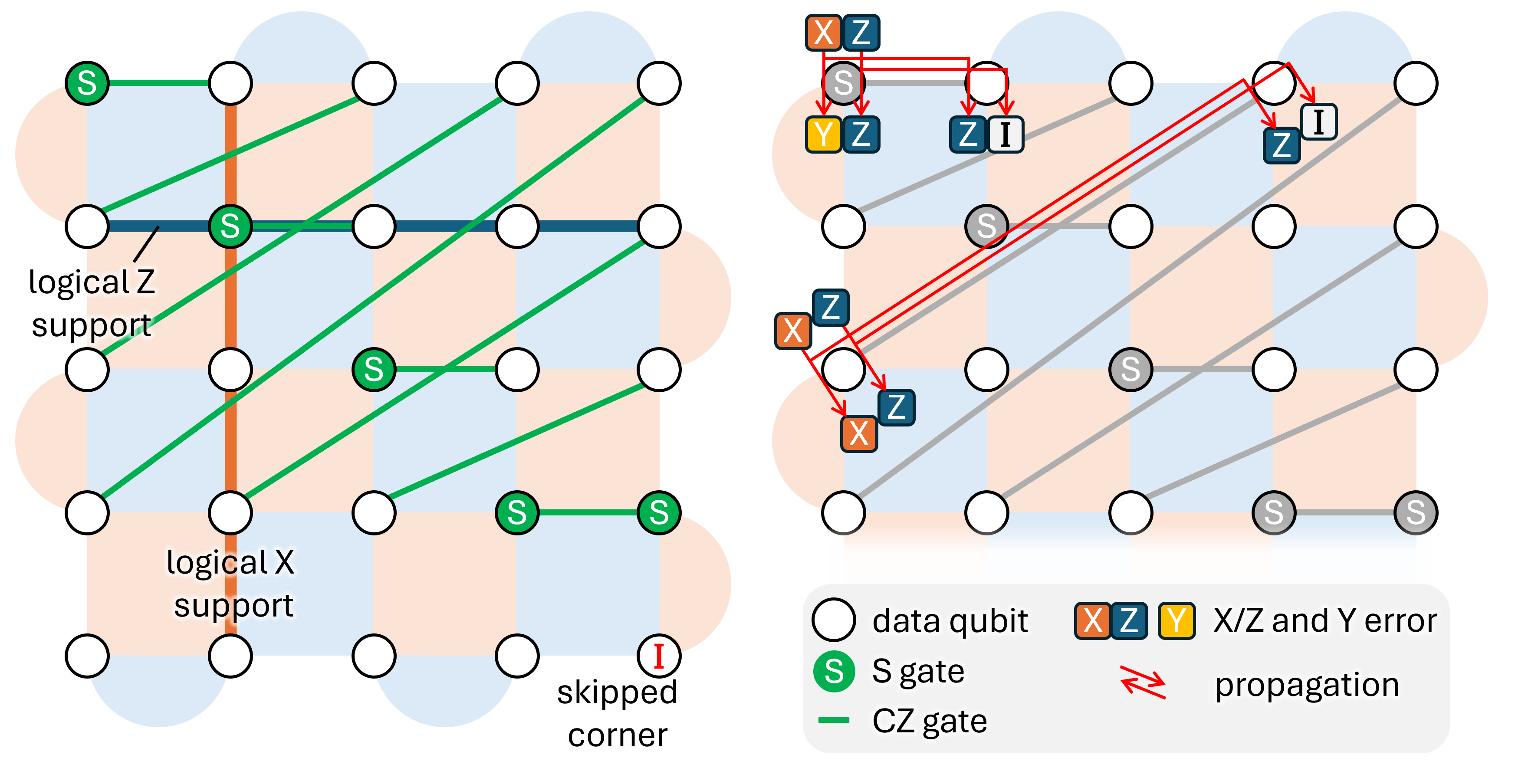}
    \caption{\textit{Left side:} $d=5$ rotated surface code patch showing the physical $S$ and $CZ$ gates used in the logical $S$ gate construction of Ref.~\cite{holmes2026quantum}.
    \textit{Right side:} The propagation logic used by IRS to convert the logical $S$ gate decoding problem into a memory-like decoding problem. Physical gates are de-saturated for readability. The propagation of a $Y$ error follows directly from $X$ and $Z$.}
    \label{fig:S-gate-implementation}
\end{columnfigure}

The propagation rule for IRS follows directly from $S$ gate conjugation.
A phase gate maps $X \mapsto Y$ and leaves $Z$ unchanged (conjugation maps $Z$ to itself), and a controlled-$Z$ maps $X$ on either qubit to $X$ on that qubit and $Z$ on its partner, while leaving $Z$ unchanged on both. 
That is, where $U\circ P = UPU^\dagger$,
\begin{gather}
    S\circ X=Y, \quad S\circ Z=Z \nonumber \\
    CZ\circ\{I\otimes X, X\otimes I, X\otimes X\} = \{Z\otimes X, X\otimes Z, Y\otimes Y\}, \nonumber \\
    CZ\circ\{I\otimes Z, Z\otimes I, Z\otimes Z\} = \{I\otimes Z, Z\otimes I, Z\otimes Z\}.
\end{gather}
Under the layer as a whole, $Z$ errors pass through unchanged, while $X$ errors acquire a $Z$ component on any qubit where a phase gate is applied, and on the reflected qubit where a controlled-$Z$ is applied.
Unlike the closed loops of tCNOTs of Section \ref{sec:iteration-dependence}, no inferred error can return to the qubit that generated it according to these $S$ propagation rules, and a single IRS update therefore suffices: a solution is found for the $X$-decoding problem, its inferred errors are converted into the propagation correction for the $Z$ problem, and the $Z$ problem is solved against the corrected record.

In Figure \ref{fig:S-gate-decoding}, we compare the decoding results for the logical $S$ gate using IRS and LOM against a naive MWPM decoder without propagation updates.
We compare the decoders on a circuit that performs repeated pairs of logical $S$ gates, specifically $d+1$ gates for odd code distances $>3$.
This is to confirm that repeated $S$ gates do not pose a difficult decoding problem, despite error propagation by the $CZ$ gates internal to the gate implementation, as well as to verify that distance protection is preserved~\footnote{Though not shown, we verified that IRS successfully decodes a single logical $S$ gate applied to a $\ket{+}_L$ state and measuring logical $Y$ up to $d=9$. This represents a less intensive test than Figure~\ref{fig:S-gate-decoding} so is not included, achieving a logical error rate of $1.46\time10^{-6}$ for a single $S$ gate followed by $Y_L$ measurement, compared to $1.89\times10^{-6}$ for two $S$ gates followed by $X_L$ measurement, using the same error model as in Figure~\ref{fig:S-gate-decoding}}. 
Both IRS and LOM outperform naive MWPM and retain distance-dependent suppression over the range shown, with IRS performing better  as $d$ increases.
At $d=9$, IRS gives a circuit logical error rate of approximately $2.5\times 10^{-5}$ with one initial and one final round, compared with approximately $10^{-4}$ for LOM, a reduction by a factor of four.
Increasing the number of SE rounds as part of state preparation and immediately before measurement to $\lceil d/2\rceil$ reduces the IRS error rate to approximately $10^{-5}$, while LOM remains close to $10^{-4}$, widening the advantage to roughly an order of magnitude.
These results show that the same propagation-assisted local approach applies to correlations between the $X$- and $Z$-decoding problems within a single logical block, rather than only to propagation between logical blocks.

\begin{columnfigure}[ht]
    \centering
    \includegraphics[width=\linewidth]{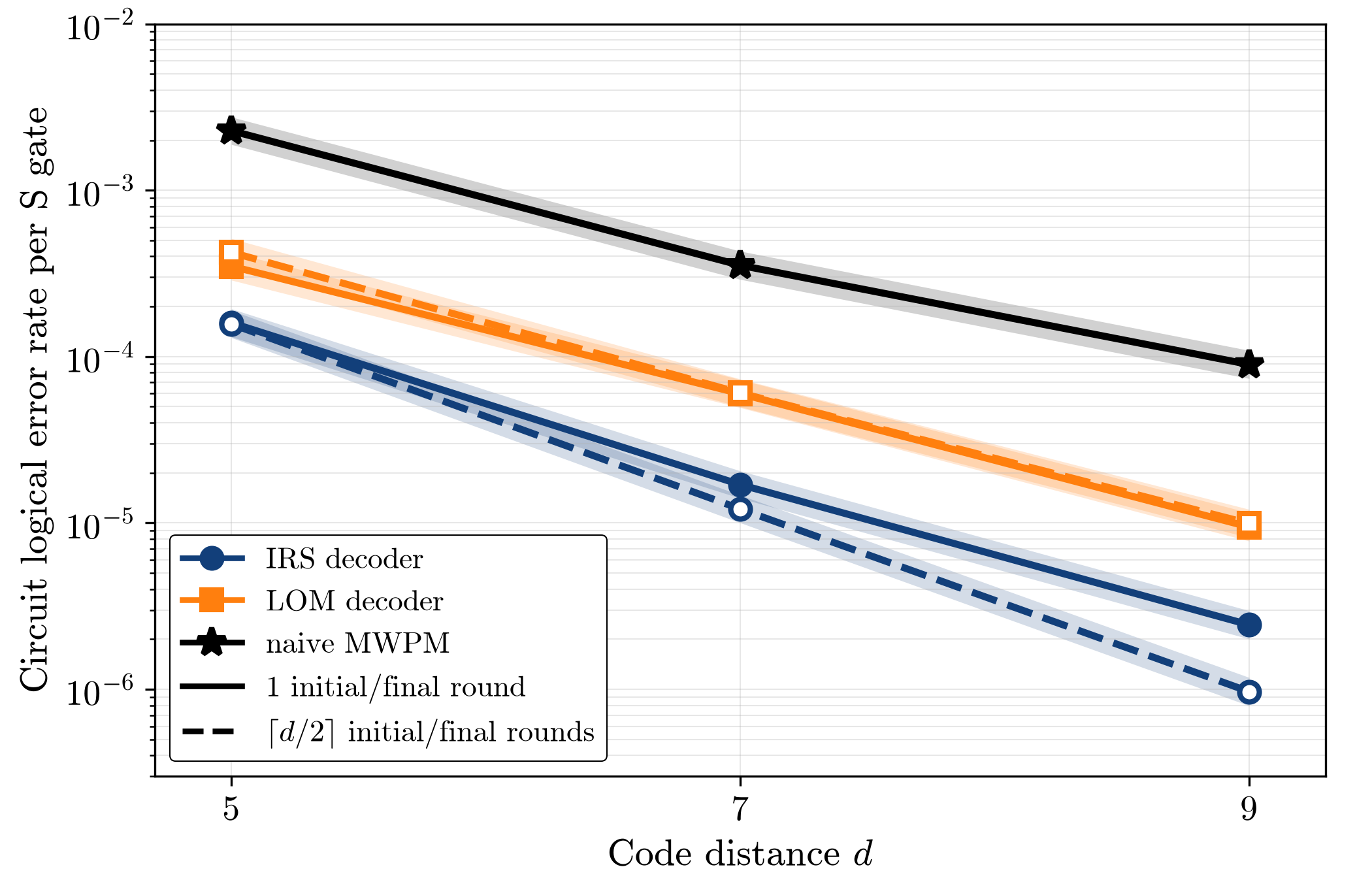}
    \caption{Logical error rate per $S$ gate with physical noise $p=10^{-3}$. Each circuit prepares $\lvert+\rangle_L$ and applies $d+1$ $S_L$ gates before measuring $X_L$, resulting in an effective circuit action of $Z,I,Z$ for $d=5,7,9$, respectively. Solid lines with filled markers, and dashed lines with empty markers, use one and $\lceil d/2\rceil$ extraction rounds before the first gate and after the last, respectively. Blue circles show IRS with correlated matching, orange squares show LOM, and black stars show naive MWPM for the one-round padding schedule. Two-qubit gates have depolarizing probability $p$; state preparation, reset, idle, and physical $S$ gates have single-qubit depolarizing probability $p/10$. Ancilla measurement outcomes flip with probability $p$ while end-state data measurement is noiseless, for consistency with the literature and Stim's~\cite{gidney2021stim} measure Pauli-product (MPP) function. Each point is estimated from at least $100$ observed logical failures, with the shaded bands indicating the 95\% confidence Wilson interval.}
    \label{fig:S-gate-decoding}
\end{columnfigure}

\section{Discussion and Conclusions}\label{sec:discussion}
\begingroup
\setlength{\parskip}{0pt}
The prospect of reducing the number of SE rounds between logical gates from $O(d)$ to $O(1)$ is a central motivation of transversal computation~\cite{cain2024correlated}.
The optimal number of rounds between tCNOT layers depends on the circuit structure, decoder, and error model. 
In this work, we have developed our previous decoding scheme~\cite{wan2024iterative}.
We retain its core propagate-and-redecode procedure using memory-like detectors and local matching with correlations accounted for by propagation.
We have added multiple new features, resulting in what we call the Iterative Redecode Scheme (IRS), benchmarked against other transversal decoders from the literature. 

\textbf{Propagation-Assisted Local (PAL) decoders.} 
We identified a class of transversal decoders in the literature, distinguishing between the core properties of different approaches. 
In PAL decoders, correlations between otherwise local decoding problems are passed between logical blocks, rather than decoded as a single global hypergraph. 
IRS uses the propagation correction to ensure consistency in the decoding solution between blocks. 

\textbf{Prior-informed DEM.} 
We recognised that the first IRS iteration decodes against a DEM without mechanisms able to explain propagated contamination. 
This biases the initial matching, and biases the propagation correction built from it. 
We dealt with this by augmenting the memory-equivalent DEM with local, graph-like mechanisms for the expected contamination across each layer, composing probabilities by parity, and staging the removal of the prior layer-by-layer through the first iteration to avoid double counting.

\textbf{Indeterminate stabilizer handling.} 
We represented preparation-induced stabilizer randomness as zero-weight physical Pauli mechanisms, which lets IRS keep memory-like detectors, infer the gauge contribution, and propagate it consistently through the circuit. 
This makes GHZ and Bell preparation tractable, from which we show that additional initialisation rounds specifically benefit IRS in a manner that cannot be leveraged by other decoders.

\textbf{Worst-case and average-case circuit analysis.} 
We have identified a particularly adversarial class of circuits where an error can propagate back to the block it originated from across the layers of a circuit.
We studied IRS decoding convergence and logical performance for alternating transversal CNOT circuits as the prototypical example of this, and found that IRS maintains logical performance at the cost of additional decoding iterations and an optimal SE buffer between tCNOTs of $\lfloor d/4\rfloor$, still well below $d$. 
Circuits which do not involve this propagation mechanism require only a single separating SE round, as illustrated by the Reed--Muller $15$-to-$1$ distillation simulations of Ref.~\cite{wan2024constant}. 
This is compared to a deep random Clifford circuit, which sits between the two; adversarial gadgets within these circuits that share the propagation property of alternating tCNOTs are identified as the limiting factor to an $O(1)$ SE buffer. 

\textbf{Finite state-space analysis for convergence.} 
We established that a complete iteration is a deterministic map on a finite state space, so an uncapped trajectory must reach a fixed point or periodic orbit, and decomposed the logical error rate by convergence bin, showing that most failures occur in rapidly converging fixed points, not in cycles.

\textbf{Cycle-detection and early termination.} 
We measured the cycling population ($\sim 0.1\%$ of shots), its logical error rate (the highest of any bin), and its contribution to the total (an order of magnitude below the first bins).
We showed that shots which have entered a cycle can be detected and terminated with negligible impact on performance, potentially enabling post-selection if needed. 
We also introduced an iteration cap and showed that iterating until convergence is rarely necessary to achieve sufficient logical error rates, with diminishing returns for iterations beyond one or two. 
In practice, IRS needs only a couple of iterations and close to $O(1)$ separating SE rounds in most cases, depending on the frequency of adversarial gadgets within the circuit. 
In combination with the prior-informed DEM, this represents a sizeable improvement compared to the original method~\cite{wan2024iterative} in computation cost and logical performance. 

\textbf{Extension to the logical $S$ gate.} 
We have decoded a recent implementation~\cite{holmes2026quantum} of the logical $S$ gate using physical $CZ$ and $S$ gates in which propagation within a block still occurs.
This demonstrates that IRS can be used to decode beyond tCNOTs.

\textbf{Benchmarking and comparisons.} 
Under matched circuit and noise assumptions, we benchmarked IRS against the ghost protocol~\cite{turner2026scalable}, using its authors' circuits and reference data supplied by them; logical observable matching (LOM)~\cite{serra2026decoding}, using the open-source implementation validated against published trends; and reliable logical Pauli-product (RLPP) decoding~\cite{cain2025fast}, which shares largely the same core method as LOM, using its published GHZ benchmark. Notably, Ref.~\cite{cain2025fast} had previously presented this benchmark as exposing a weakness of the earlier iterative scheme~\cite{wan2024iterative}.
In every case, IRS outperforms the available decoders from the literature, compared with published results in the case of the ghost protocol and RLPP (where open source implementations are not available), or simulated directly for LOM (which is open source). 
Most notably, at $d=11$ we achieve a factor $4\times$ improvement in logical error rate over the ghost protocol in the deep random Clifford benchmark (Figure \ref{fig:deep-random}), and a factor $15\times$ improvement over LOM, using the same SE round buffer, and on the GHZ preparation circuit, IRS achieves half the error rate of the published RLPP results~\cite{cain2025fast}, reaching a factor $25\times$ improvement with the standard $\lceil d/2\rceil$ initialisation rounds (Figure \ref{fig:quera_ghz_graph}).

We note that a detector record can admit multiple locally preferred, globally consistent error configurations with different logical predictions.
IRS returns one without comparing likelihoods; a selection rule over this degenerate set would improve accuracy without abandoning local decoding. 
Runtime and memory use, and their scaling with code distance, circuit width and depth, remain unquantified, with real-time decoding requiring a windowed approach~\cite{wan2024iterative} to carry forward propagated information and unresolved error information consistently across window boundaries.
These are left to future work.

\begin{appendices}
\renewcommand{\thesubsection}{\thesection.\arabic{subsection}}

\section{IRS Algorithm}\label{app:algorithm}
Algorithm \ref{alg:IRS} summarizes the IRS procedure.
The memory-equivalent DEM $\mathcal{M}_b$ of each logical block $b$ is built by $\mathrm{MemoryDEM}$ from the logical preparation basis $\beta_b$. It fixes the first-round detectors and, for indeterminate initial stabilizers, supplies the zero-weight mechanisms of Sec. \ref{sec:initial-stabilizers}.
From the physical error model $\mathcal{P}$, the function $\mathrm{PropagatePrior}$ returns $\mathcal{Q}_{b,\ell}$, the local prior mechanisms contributed to block $b$ by tCNOT layer $\ell$, constructed as described in Sec \ref{sec:calibration}.
The DEM decoded against, $\mathcal{D}_b$, begins as the prior-informed DEM $\mathcal{M}_b\cup\bigcup_\ell\mathcal{Q}_{b,\ell}$, and each $\mathcal{Q}_{b,\ell}$ is removed once layer $\ell$ has been explicitly corrected, so that by the end of the first iteration $\mathcal{D}_b=\mathcal{M}_b$ and remains so for every subsequent iteration.

IRS begins with an initial MWPM decode of the measured detector record $\mathbf{s}$ against $\{\mathcal{D}_b\}$.
The circuit is divided into ordered tCNOT layers $G_\ell$, with $(C,T)\in G_\ell$ denoting a tCNOT from control block $C$ to target block $T$ at time $t_\ell$.
For each layer, $\boldsymbol{\delta}_\ell$ is the binary detector-toggle vector of the current propagation correction.
These layer corrections persist between iterations, necessitating that when layer $\ell$ is revisited, its previous correction is first removed from $\mathbf{s}$ before an updated correction is inferred and applied.
The quantities $f_b^X$ and $f_b^Z$ denote the propagated $X$- and $Z$-sector Pauli frames accumulated on logical block $b$ from earlier tCNOT layers within the current iteration, and are reset at the start of each iteration.

For a tCNOT $(C,T)$ in layer $\ell$, $\mathcal{E}_{C,\ell}^{X}$ and $\mathcal{E}_{T,\ell}^{Z}$ denote the $X$ error on the control and $Z$ error on the target respectively, present immediately before the gate, as inferred from the current MWPM solution, including the matching-edge correction of Eq. \eqref{eq:tcnot_det_transform_with_correction}.
The quantities $\pi_X$ and $\pi_Z$ are the corresponding Pauli-error contributions propagated across that tCNOT, from control to target and target to control, respectively.
The function $\mathrm{OddParityDetectors}(\pi_X,\pi_Z)$ maps these propagated Pauli errors onto the propagation correction $\boldsymbol{\delta}_\ell$ of the receiving blocks.
After each tCNOT layer, the modified detector record is immediately re-decoded so that the estimate used at layer $\ell+1$ incorporates all propagation corrections inferred at preceding layers in the same iteration.
The complete propagation state after iteration $n$ is denoted $\boldsymbol{\Delta}^{(n)}=(\boldsymbol{\delta}_1,\ldots,\boldsymbol{\delta}_L)$, and $\mathcal{H}$ collects the states of all preceding iterations. 

Three possible conditions end the loop, recorded in $\mathrm{\scriptstyle STATUS}$. 
If $\boldsymbol{\Delta}^{(n)}=\boldsymbol{\Delta}^{(n-1)}$ the propagation state is unchanged and IRS has reached a fixed point.
Otherwise, if $\boldsymbol{\Delta}^{(n)}\in\mathcal{H}$, an earlier state has recurred; since the update rule is deterministic, the same sequence must repeat indefinitely and a cycle is flagged. 
The fixed-point test is applied first, so the membership test is effectively against $\{\boldsymbol{\Delta}^{(0)},\ldots,\boldsymbol{\Delta}^{(n-2)}\}$. 
A fixed point is a period-one cycle, but a fixed point and longer cycles are reported separately because they have opposite implications for the decoding solution. 
If neither condition is met, the loop runs to the cap $N$ and the current estimate is returned. 
IRS returns $\mathrm{\scriptstyle STATUS}$ and the terminating index $n$ alongside the error estimate; these are the quantities by which shots are binned in Sec. \ref{sec:iteration-dependence}. 

\begin{algorithm}[H]
\caption{Iterative Redecode Scheme (IRS) decoder.}
\label{alg:IRS}

\begin{algorithmic}[1]

\Require Measured detector record $\mathbf{s}$
\Require Logical preparation basis $\beta_b$ for each logical block $b$
\Require Physical error model $\mathcal{P}$
\Require Ordered tCNOT layers $\{G_\ell\}_{\ell=1}^{L}$
\Require Maximum number of iterations $N$

\State $\{\mathcal{M}_b\} \gets \mathrm{MemoryDEM}\!\left(\{\beta_b\}\right)$
\Comment{initialisation bookkeeping, Sec. \ref{sec:initial-stabilizers}}

\State $\mathcal{Q}_{b,\ell} \gets \mathrm{PropagatePrior}\!\left(\mathcal{P},G_\ell\right)$ \textbf{ for all } $b,\ell$
\Comment{layer-$\ell$ prior mechanisms on block $b$}

\State $\mathcal{D}_b \gets \mathcal{M}_b\cup\bigcup_{\ell=1}^{L}\mathcal{Q}_{b,\ell}$ \textbf{ for all } $b$
\Comment{prior-informed DEM, Sec. \ref{sec:calibration}}

\State $\hat{\mathbf e} \gets \mathrm{MWPM}\!\left(\mathbf{s},\{\mathcal{D}_b\}\right)$
\Comment{initial local decode}

\State $\boldsymbol{\delta}_\ell \gets \mathbf{0}$ \textbf{ for all } $\ell$
\Comment{stored detector toggles for each layer}

\State $\mathcal{H} \gets \varnothing$
\Comment{propagation states already visited}

\State $\mathrm{\scriptstyle STATUS} \gets \textsc{cap}$
\Comment{overwritten if the loop terminates early}

\For{$n=0,\ldots,N-1$}
\Comment{iteration cap $N$}

    \ForAll{logical blocks $b$}
        \State $f_b^X \gets \mathbf{0}$, $f_b^Z \gets \mathbf{0}$
        \Comment{reset propagated frame each iteration}
    \EndFor

    \For{$\ell=1,\ldots,L$}

        \State $\mathbf{s} \gets \mathbf{s}\oplus\boldsymbol{\delta}_\ell$
        \Comment{remove this layer's correction from the previous iteration}

        \State $\boldsymbol{\delta}_\ell \gets \mathbf{0}$

        \ForAll{tCNOTs $(C,T)\in G_\ell$}
            \State $\mathcal{E}_{C,\ell}^{X} \gets \mathrm{InferX}\!\left( \hat{\mathbf e},C,t_\ell \right)$

            \State $\mathcal{E}_{T,\ell}^{Z} \gets \mathrm{InferZ}\!\left( \hat{\mathbf e},T,t_\ell \right)$

            \State $\pi_X \gets \mathrm{Transport}\!\left( \mathcal{E}_{C,\ell}^{X}\oplus f_C^X,\, C\rightarrow T \right)$

            \State $\pi_Z \gets \mathrm{Transport}\!\left( \mathcal{E}_{T,\ell}^{Z}\oplus f_T^Z,\, T\rightarrow C \right)$

            \State $f_T^X \gets f_T^X\oplus\pi_X$
            
            \State $f_C^Z \gets f_C^Z\oplus\pi_Z$

            \State $\boldsymbol{\delta}_\ell \gets \boldsymbol{\delta}_\ell \oplus \mathrm{OddParityDetectors}(\pi_X,\pi_Z)$

        \EndFor

        \State $\mathbf{s} \gets \mathbf{s}\oplus\boldsymbol{\delta}_\ell$
        \Comment{apply updated correction}

        \State $\mathcal{D}_b \gets \mathcal{D}_b\setminus\mathcal{Q}_{b,\ell}$ \textbf{ for all } $b$
        \Comment{stage away this layer's prior; no effect after the first iteration}

        \State $\hat{\mathbf e} \gets \mathrm{MWPM}\!\left( \mathbf{s},\{\mathcal{D}_b\} \right)$
        \Comment{re-decode after each layer}

    \EndFor

    \State $\boldsymbol{\Delta}^{(n)} \gets \left( \boldsymbol{\delta}_1,\ldots,\boldsymbol{\delta}_L \right)$

    \If{$n>0$ \textbf{and} $\boldsymbol{\Delta}^{(n)} = \boldsymbol{\Delta}^{(n-1)}$}
        \State $\mathrm{\scriptstyle STATUS} \gets \textsc{fixed point}$
        \State \textbf{break}

    \ElsIf{$\boldsymbol{\Delta}^{(n)} \in \mathcal{H}$}
        \State $\mathrm{\scriptstyle STATUS} \gets \textsc{cycle}$
        \Comment{an earlier state has recurred; no fixed point will be reached}
        \State \textbf{break}
    \EndIf

    \State $\mathcal{H} \gets
    \mathcal{H}\cup\{\boldsymbol{\Delta}^{(n)}\}$

\EndFor

\State $\mathbf f \gets \{f_b^X,f_b^Z\}_b$
\State \Return $\hat{\mathbf e}$ composed with the propagated frame $\mathbf f$, together with $\mathrm{\scriptstyle STATUS}$ and the terminating iteration index $n$

\end{algorithmic}
\end{algorithm}

\section{Stabilizer frame used by IRS}

\subsection{Stabilizer transport and detector structure across a tCNOT}\label{app:stabilizer-transport}
A stabilizer code is defined by a set of mutually commuting Pauli operators $S_i$, named stabilizers, whose common $+1$ eigenspace defines the codespace. 
Measuring a stabilizer generator $S_i$ gives an eigenvalue $\pm 1$, which we represent as a binary measurement outcome. 
A Pauli error which anticommutes with $S_i$ flips this outcome. 
During a memory experiment, detector events are formed by comparing repeated measurements of the same stabilizer between syndrome-extraction rounds, such that a change in the measured stabilizer value produces a detector event.

This picture changes in the presence of transversal logical gates because the stabilizers themselves are transported through the logical circuit. 
For a Clifford operation $U$, a stabilizer $S_i$ transforms under conjugation as $S_i\mapsto U S_i U^\dagger$. 
For a tCNOT, this couples stabilizers on the control and target logical blocks: $Z$-type stabilizers, $Z_i$, propagate from control to target, while $X$-type stabilizers, $X_i$, propagate from target to control. 
Figure \ref{fig:propagated_z} illustrates this propagation for the $Z$-type stabilizers. 
Using $Z_i^B(t)$ and $X_i^B(t)$ to denote the binary measurement outcomes of the corresponding stabilizers on block $B\in{C,T}$, where $C$ and $T$ denote the control and target logical blocks, at time $t$, a tCNOT applied between measurements at $t$ and $t+1$ gives
\begin{align}\label{eq:tcnot_stab_transform}
    Z_{i}^{T}(t+1) \longmapsto& Z_{i}^{T}(t+1) \oplus Z_{i}^{C}(t) \nonumber \\
    X_{i}^{C}(t+1) \longmapsto& X_{i}^{C}(t+1) \oplus X_{i}^{T}(t),
\end{align}
where $i$ labels the corresponding ancilla qubit responsible for the stabilizer measurement.

\begin{columnfigure}[ht]
    \centering
    \includegraphics[width=\linewidth]{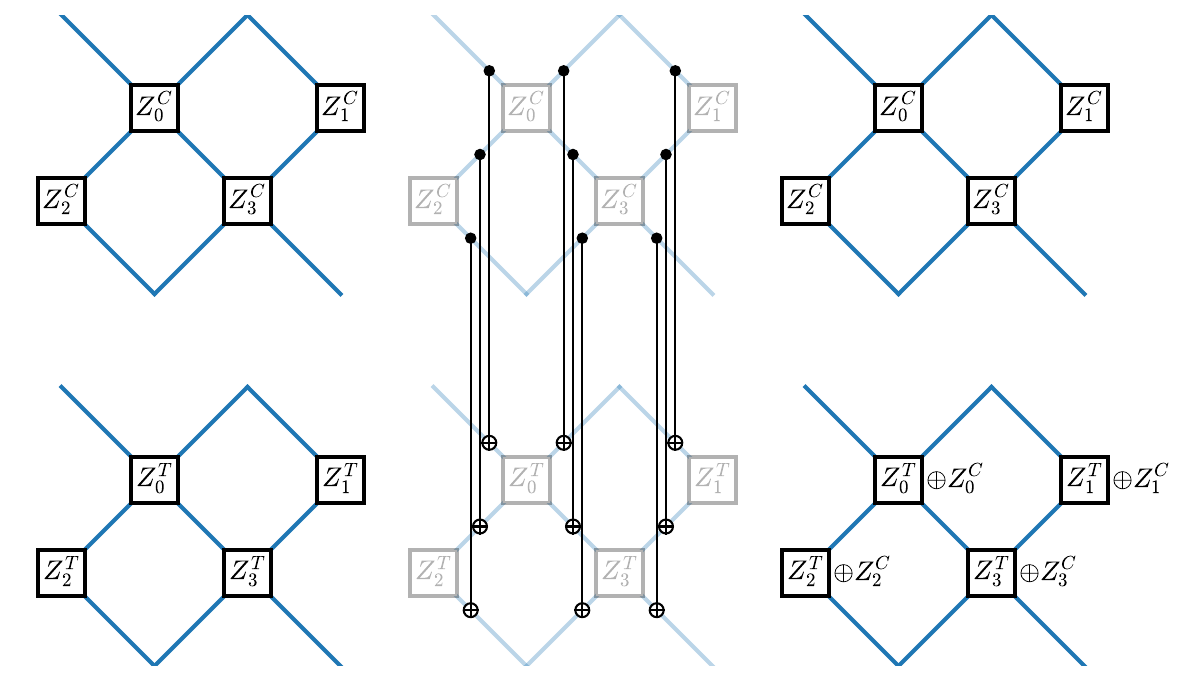}
    \caption{Schematic of a distance-3 surface-code before (\textit{left}, with stabilizer measurements at $t$), during (\textit{middle}), and after (\textit{right}, with stabilizer measurements at $t+1$) a tCNOT, as time runs left to right. Edges represent data-qubits while intersections (square boxes) represent ancilla-qubits that store the $Z$-parities of their neighbouring data-qubits. The tCNOT XORs the $Z$-parities of the control data-qubits to those of the target data-qubits.}
    \label{fig:propagated_z}
\end{columnfigure}

Rather than redefining the detectors to include this stabilizer transport, IRS deliberately retains the same memory-like detector basis across the tCNOT.
The transported contribution therefore appears directly in the local detector record. 
A detector event on one block can consequently arise either from a fault generated locally on that block or from an error propagated from another block through the tCNOT. 
From the perspective of the memory-equivalent DEM, the latter appears as tCNOT-induced contamination because the corresponding cross-block fault mechanism is absent from the local memory error model; this creates two further related problems for independent block decoding. 
\begin{enumerate}
    \item Each logical block contains both locally generated and propagated errors. 
    The propagated contribution therefore increases the effective error burden seen by the local decoder, and can combine with locally generated errors to form an undetected logical error. 
    In deep circuits, where errors may be propagated through multiple tCNOT layers, neglecting this contribution can degrade both the threshold and sub-threshold logical-error scaling.
    \item Independent local block decoders also discard the information that detector events on different blocks may originate from the same underlying fault. 
    A propagated fault is instead assigned separate local explanations on the affected blocks, so its likelihood can be effectively double-counted and the inferred corrections become systematically biased. 
    These incorrect local error estimates may then themselves be propagated through subsequent tCNOT layers.
\end{enumerate}

IRS addresses these effects by iteratively estimating the propagated contribution to each local detector record and removing it before memory-like decoding. 
The removed contribution is tracked separately, leaving a residual detector record which is, as closely as possible, generated only by locally occurring faults and can therefore be decoded using the memory-equivalent DEM. 
The inferred local errors and propagated contributions are subsequently recombined when constructing the final error and logical-observable prediction. 
The following subsection develops this propagation correction explicitly.

\subsection{Recovery of Memory stabilizers and Propagated decoding}\label{app:memory-stabilizers}
The first step in the propagation correction is to determine the detector contribution that should be propagated across each tCNOT. 
Since the CNOT transformation in Eq. \eqref{eq:tcnot_stab_transform} is self-inverse, the transported stabilizer contribution can be removed by applying the same transformation to the measured detector record.

The memory-like parity detectors are defined by
\begin{align}
    D_{P,i}(0) &= P_i(0), \qquad D_{P,i}(t) = P_i(t)\oplus P_i(t-1), 
\end{align}
where the subscript $P \in \{X, Z\}$ denotes the type of stabilizer, $i$ denotes the index of the stabilizer at a given time $t$ and, $P_i(t)$ is the measured value of the stabilizer. Solving the recursive equation for $P_i(t)$ yields,
\begin{equation}\label{eq:measurement_from_detectors}
    P_i(t)=\bigoplus_{r=0}^{t}D_{P,i}(r).
\end{equation}
Substituting these relations into Eq. \eqref{eq:tcnot_stab_transform} (where a tCNOT is applied between times $t$ and $t+1$) gives the corresponding transformation directly in terms of the detector record,
\begin{subequations}
\label{eq:tcnot_det_transform}
\begin{align}
    D_{Z,i}^{T}(t+1) &\longmapsto D_{Z,i}^{T}(t+1) \oplus \bigoplus_{r=0}^{t}D_{Z,i}^{C}(r), \label{eq:tcnot_det_transform_z} \\
    D_{X,i}^{C}(t+1) &\longmapsto D_{X,i}^{C}(t+1) \oplus \bigoplus_{r=0}^{t}D_{X,i}^{T}(r). \label{eq:tcnot_det_transform_x}
\end{align}
\end{subequations}
We refer to these detector toggles as the \textit{propagation correction}. 
They remove the contribution predicted from stabilizer transport through the tCNOT from the local detector record.

The cumulative detector history in Eq. \eqref{eq:measurement_from_detectors}, reconstructs the \emph{measured} stabilizer value. 
In the presence of measurement errors or data errors whose matching edges span the tCNOT, this need not give the data-error configuration which should be propagated through the gate. 
For example, a measurement error immediately before the tCNOT can produce one detector event before the gate and a second after it. 
Directly applying Eq. \eqref{eq:tcnot_det_transform} would treat the first event as evidence for an error already present before the tCNOT and would therefore propagate a contribution which should instead be attributed to the time-spanning measurement error.

IRS resolves this ambiguity using the local MWPM solution. 
Each logical block is decoded against its memory-equivalent DEM, and matching edges which cross the tCNOT boundary identify error mechanisms which have not been resolved within the detector history up to time $t$. 
These edges provide an additional correction to the reconstructed stabilizer value. Eq. \eqref{eq:tcnot_det_transform_z} therefore becomes
\begin{align}\label{eq:tcnot_det_transform_with_correction}
    D_{Z,i}^{T}(t+1) \longmapsto& D_{Z,i}^{T}(t+1) \oplus \bigoplus_{r=0}^{t}D_{Z,i}^{C}(r) \nonumber \\
    &\quad \oplus
    \underbrace{\bigoplus_{s=t+1}^{\infty} \alpha_{Z,i}(s)D_{Z,i}^{C}(s)}_{\text{matching-edge correction}}, 
\end{align}
where $\alpha_{P,i}(s)=1~\forall P\in\{X,Z\}$,  if the MWPM solution contains an edge matching a detector $D_{P,j}^{C}(r)$ at $r\leq t$ to $D_{P,i}^{C}(s)$ at $s>t$, and is zero otherwise. 
Eq.~\eqref{eq:tcnot_det_transform_x} is corrected similarly.

As a simple example, consider a measurement error on $Z_i^C(t)$ immediately before the tCNOT, which produces detector events $D_{Z,i}^C(t)$ and $D_{Z,i}^C(t+1)$. 
If MWPM correctly identifies the time-like edge of this measurement error across the tCNOT, the detector event at time $t$ should not contribute to the inferred data error present before the gate. 
The final term in Eq. \eqref{eq:tcnot_det_transform_with_correction} achieves this by removing this contribution from the stabilizer value used in the propagation correction. 
Figure \ref{fig:error-prop} illustrates the same construction for both time-like and diagonal space-time matching edges which span the tCNOT.
This construction is equivalent to the Pauli-frame propagation used in Ref.~\cite{wan2024iterative} where the cumulative detector history accounts for the inferred error configuration before the tCNOT, while matching edges spanning the tCNOT boundary correct for error mechanisms which are not resolved within that history. 
The resulting estimate determines which detector events should be propagated to the neighbouring logical block.

\begin{widefigure}[ht]
    \centering
    \subfloat[\label{fig:placeholder1}]{
        \includegraphics[width=.48\linewidth]{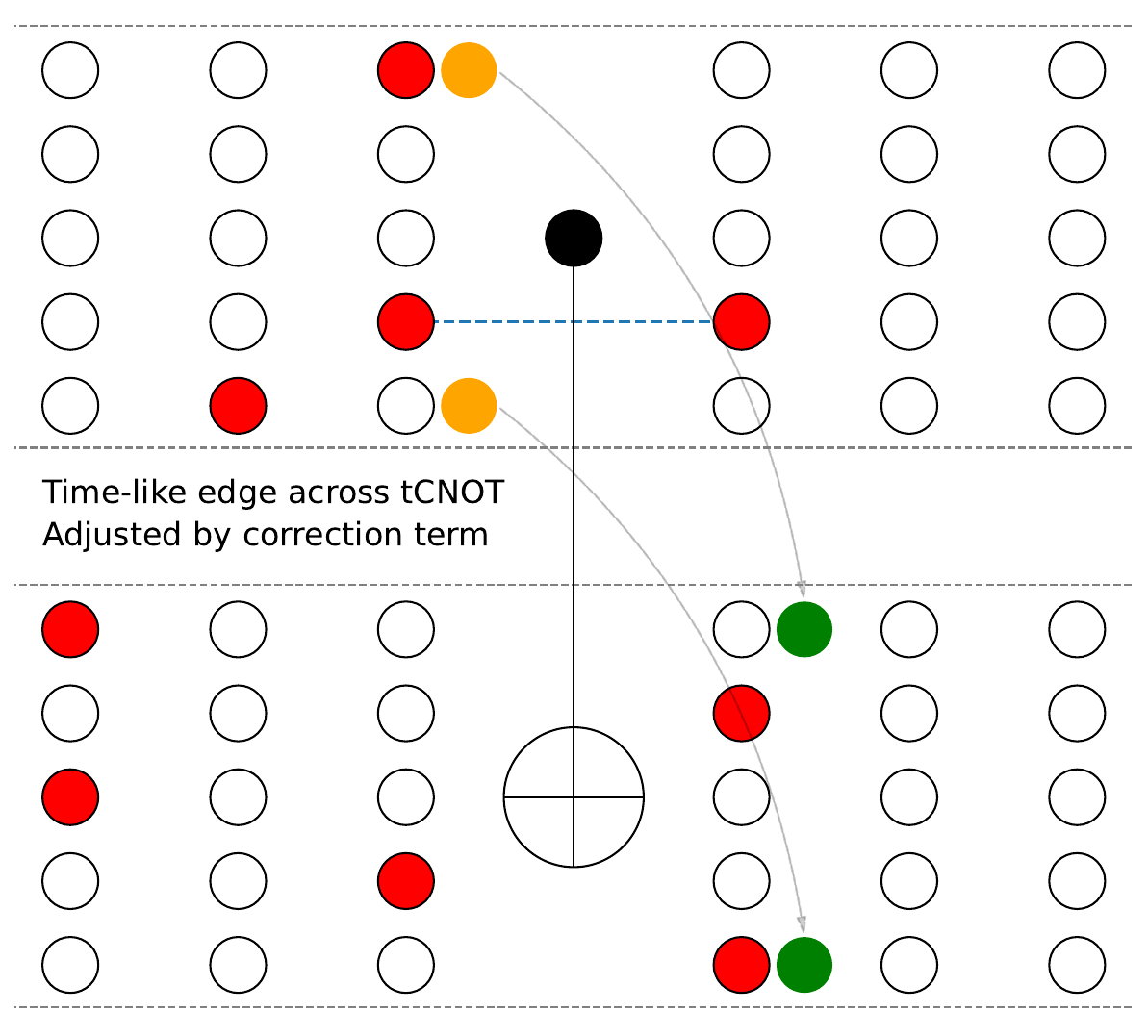}
    }
    \hfill
    \subfloat[\label{fig:placeholder2}]{
        \includegraphics[width=.48\linewidth]{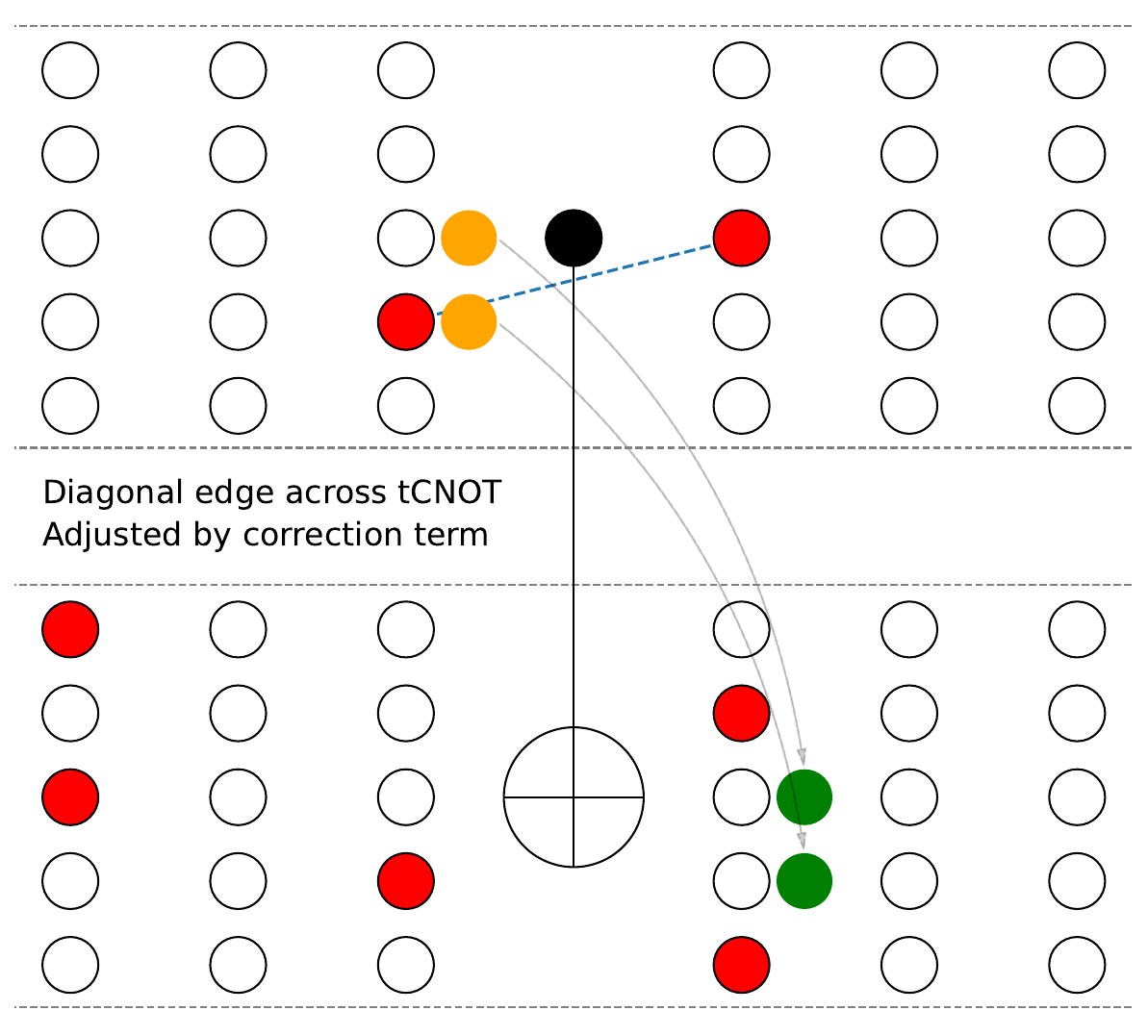}
    }
    \caption{Schematic representation of $X$-flip detectors-events being propagated across a tCNOT from  control to target. Columns of circles represent detectors at a particular time step, with red circles denoting that a detector is triggered. Before the tCNOT, the sum of detector values (the second term in Eq. \eqref{eq:tcnot_det_transform_with_correction}) is denoted by an orange circle if the value is 1. The propagated syndromes that toggle detectors on the target qubit after the tCNOT are denoted by green circles. 
    (a) Dotted-lines match two detectors across the tCNOT that would be matched with time-like edge by MWPM. This is handled by the third term in Eq. \eqref{eq:tcnot_det_transform_with_correction}, which acts as a correction. 
    (b) Dotted-lines match two detectors across the tCNOT with a diagonal space-time edge. The correction term (Eq. \eqref{eq:tcnot_det_transform_with_correction}) applies to lit detectors after the tCNOT control, propagating them as if they occured before the tCNOT onto the target patch.}
    \label{fig:error-prop}
\end{widefigure}

\section{Constructing a prior-informed DEM}\label{app:prior-informed}
The purpose of the \emph{prior-informed DEM} is to give the local decoder an appropriately weighted \emph{approximate} explanation of detector events which are expected to appear through propagation before those correlations have been explicitly handled by a complete IRS iteration.
The inclusion of this prior information adds local graph-like error mechanisms at the detector locations where tCNOT-induced contamination is expected to appear, causing a re-weighting of the existing edges in the DEM to account for these detector events.

The construction has two parts: a forward pass that advances the $X$- and $Z$-sector marginals of every data qubit through the circuit, and the formation of the DEM mechanisms that this pass implies. 
For a simple marginal construction where we track the marginal probabilities of $X$ and $Z$ components independently,  we maintain each data qubit's current probability that its supporting stabilizer measurements will be $-1$, and whether this results in a triggered detector. 
This means propagating the distribution of errors that have occurred up to the current layer through the current layer, applying appropriate commutation relations such as a Hadamard converting an observed $Z$ error into an observed $X$ error, and combining it with the distribution of errors that might then occur after that layer. 
Contributions of the same type acquired on the same data qubit cannot simply be added since two identical Pauli components cancel. 
Every probability used to build the prior is therefore not the raw accumulated probability that a mechanism causes a detector to fire, but the probability that an odd number of independent contributions is present, since contributions of the same Pauli type on the same data qubit cancel in pairs.

For each Pauli-sector $X$ and $Z$, define the bit $a_i \in \{0,1\}$ as equal to $1$ when an error of that type would be observed on data qubit $i$.
Having marginalised the distribution down to the probabilities of $X$ and $Z$, rather than tracking the full $\{I,X,Y,Z\}$ distribution, observing an error in the $X$ or $Z$ sector includes the effect of physical $Y$ errors, and commutation of the error terms with the actual circuit gates can transform between sectors; we will refer what is observed in each sector as the \textit{contribution} to the raw error model, since it combines both the impact of marginalisation with commuting the error model through the circuit layer. 
At a given layer of the circuit, we are interested in quantity $\Pr[\bigoplus_i a_i = 1]$, which is the probability that an odd number of errors of the same type cause a supporting stabilizer measurement of $-1$.
Replacing each bit by the sign variable $s_i = (-1)^{a_i} \in \{-1,+1\}$ turns the exclusive-or into a product,
\begin{equation}\label{eq:bits_to_signs}
    (-1)^{a_1 \oplus \cdots \oplus a_k} = \prod_{i=1}^{k} (-1)^{a_i} = \prod_{i=1}^k s_i.
\end{equation}
This is nothing more than a change of variables to exploit the fact that the composition by parity of bits becomes simple multiplication of the corresponding signs.
The expectation value of a single sign variable is given by
\begin{align}
    \mathbb{E}[s_i] &= (+1)(1-\mathbb{E}[a_i])+(-1)\mathbb{E}[a_i] = 1-2\mathbb{E}[a_i] \label{eq:expectation} \\
    &= 1-2p_i,
\end{align}
since for a Bernoulli bit we have 
\begin{align}
    \mathbb{E}[a_i]&=0\cdot\Pr[a_i=0]+1\cdot\Pr[a_1=1]=\Pr[a_i=1] \\
    &\equiv p_i,
\end{align}
for independent contributions $i$. 
Writing the overall sign as
\begin{equation}\label{eq:S}
    S = \prod_i s_i,
\end{equation}
Eq. \eqref{eq:bits_to_signs} gives $S = -1$ precisely when $\bigoplus_i a_i = 1$, so $S$ is itself a sign variable of the form treated in Eq. \eqref{eq:expectation}. 
Applying Eq. \eqref{eq:expectation} to $S$, and separately using the independence of the individual $i$ sign variables and the fact that the expectation of the product is the product of the expectations, gives
\begin{equation}
    1 - 2\Pr\!\left[\bigoplus_i a_i = 1\right] = \mathbb{E}[S] = \prod_{i=1}^{k} \mathbb{E}[s_i] = \prod_{i=1}^{k} (1 - 2p_i).
\end{equation}
Rearranging this allows us to define the probability that the parity of $k$ contributions will be odd as $\mathrm{PAR}(p_1, \dots, p_k)$, given by
\begin{equation}\label{eq:PAR_general}
    \mathrm{PAR}(p_1, \dots, p_k) = \frac{1}{2}\left( 1 - \prod_{i=1}^{k} (1 - 2p_i) \right).
\end{equation}

Two properties follow immediately. 
Composition is associative and commutative, so contributions may be accumulated in any order. 
And since $|1-2p_i| \leq 1$, repeated composition drives the product towards zero and $\mathrm{PAR}$ towards $1/2$ rather than towards unity: a data qubit subject to many independent Paulis becomes maximally uninformative, not certainly flipped.

For the $X$-/$Z$-decoding problem, the target/control block can acquire additional $X$-/$Z$-sector contributions, either through propagation from the control/target block, or directly from noise on the physical tCNOT. 
An $X$/$Z$ error on a data qubit is detected by the neighbouring $Z$-type/$X$-type stabilizers, so the corresponding contribution must be represented by an edge between those detectors in the syndrome-extraction round immediately following the tCNOT. 
If an edge is already present, then the weight of the existing edge is updated accordingly, otherwise a new edge is added.
If the data qubit lies on a boundary and is supported by only one relevant stabilizer, the edge instead terminates at the boundary.
In both cases, the added mechanisms remain entirely within a single logical block. 
No cross-block hyperedges are introduced, the decoding graph remains graph-like, and the same MWPM procedure used for the memory-equivalent model applies unchanged.

For each data qubit, we calculate the layer-by-layer marginal probability of carrying an $X$- and $Z$-sector contribution.
For example, consider the probability that a data qubit picks up an $X$-type error under the two-qubit depolarizing noise associated with a CNOT gate of strength $p_2$: of the fifteen non-identity two-qubit Paulis, four contain an $X$ on the specified qubit and four contain a $Y$, so the probability of acquiring an $X$-type component is therefore 
\begin{equation}\label{eq:pX}
    p_{X}=\frac{8p_2}{15}.
\end{equation}
If the data qubit participates in $n$ such gates during a syndrome-extraction round, Eq. \eqref{eq:PAR_general} with $n$ equal arguments gives its accumulated contribution as
\begin{equation}\label{eq:PARX}
    \mathrm{PAR}(p_{X},\ldots,p_{X}) = \frac{1}{2}\left(1-\left(1-\frac{16p_2}{15}\right)^n\right),
\end{equation}
with the analogous expression for the $Z$-type component. 

At each tCNOT, these marginals are advanced through the layer by composing together the contributions which propagate onto that qubit. 
The target's $X$-type probability becomes $\mathrm{PAR}$ of: its existing value, the value imported from the control, and the direct contribution from noise on the physical tCNOT.
The control's $Z$-type probability is updated correspondingly from the target. 
The control's $X$-type and the target's $Z$-type probabilities receive only the direct contribution, since nothing is imported into those sectors. 
Repeating through the remaining layers gives the expected local contamination at every stage of the circuit.

For a single data qubit, one Pauli sector, and a single tCNOT layer, there are four probabilities to consider.
Before the gate, $p_{\mathrm{acc}}$ is the probability that the qubit already had the relevant component, accumulated over every preceding layer and syndrome-extraction round since its reset .
During the gate, $p_{\mathrm{prop}}$ is the probability that a component is propagated across the gate from the partner block, and $p_{\mathrm{dir}}$ is the probability that a component is deposited by noise on the physical tCNOT itself.
After the gate, $p_{\mathrm{out}}$ is the resulting probability that, given $p_{\mathrm{acc}}$, $p_{\mathrm{prop}}$ and $p_{\mathrm{dir}}$, the qubit now carries the component immediately after the gate.

The expectation of the corresponding sign variables follows from Eq. \eqref{eq:expectation}, using the fact that the expectation of a product of independent quantities is the product of their expectations, so that the expectation of the sign variables after a tCNOT is 
\begin{equation}
    \mathbb{E}[s_\mathrm{out}] = \mathbb{E}[s_\mathrm{acc}] \mathbb{E}[s_\mathrm{prop}] \mathbb{E}[s_\mathrm{dir}], 
\end{equation}
which is equivalent to 
\begin{align}\label{eq:fourway}
    1-2p_\mathrm{out} = (1-2p_\mathrm{acc}) (1-2p_\mathrm{prop}) (1-2p_\mathrm{dir}).
\end{align}
The goal is to identify a mechanism $q$ which reproduces $p_{\mathrm{out}}$ when composed with $p_{\mathrm{acc}}$, 
\begin{equation}\label{eq:qdef}
    1 - 2p_{\mathrm{out}} = (1 - 2p_{\mathrm{acc}})(1 - 2q).
\end{equation}
Equating Eq. \eqref{eq:fourway} with Eq. \eqref{eq:qdef} gives
\begin{equation}
    1 - 2q = (1 - 2p_{\mathrm{prop}})(1 - 2p_{\mathrm{dir}}),
\end{equation}
from which it follows that 
\begin{equation}
    q = \mathrm{PAR}(p_{\mathrm{prop}}, p_{\mathrm{dir}}),
\end{equation}
which is Eq. \eqref{eq:PAR} of the main text. 
The errors already accumulated before the gate cancels identically, so the added mechanism is solely responsible for the effect of the CNOT gate itself. 
Assigning $p_{\mathrm{out}}$ to the edge directly would instead double count the history that the memory-equivalent DEM already describes through the mechanisms of earlier rounds.

For a circuit containing $L$ tCNOT layers, during the first iteration before any layer has been processed, the detector record contains contamination from all $L$ layers, so the prior-informed DEM will need to contain prior contributions associated with layers $(0,\ldots,L-1)$. 
Once layer $k$ has been processed, IRS has explicitly inferred and applied its propagation correction to the detector record. 
Retaining the corresponding prior contribution beyond this point would count the same contamination both in the corrected syndrome and in the decoding model. 
The prior associated with layer $k$ is therefore removed before the next decoding stage so that, for any stage $k$, the DEM contains prior contributions only from tCNOT layers $(k,\ldots,L-1)$.
By the end of the first iteration all tCNOT layers have been processed, all corresponding prior contributions have been removed, and the decoder has returned to the unaugmented memory-equivalent DEM. 
All subsequent IRS iterations use this DEM throughout. 

The construction is summarized by the following procedure, which depends only on the circuit and the physical error model and is therefore carried out once, before any syndrome is decoded.
It is useful to index the logical blocks by $b$, and the tCNOT layers by $\ell = 0, \dots, L-1$ in circuit order. 
We write $M_b$ for the memory-equivalent DEM of block $b$, obtained by removing the tCNOTs and treating the block as an independent memory, and $Q_{b,\ell}$ for the set of prior mechanisms that layer $\ell$ contributes to block $b$; this notation is consistent with that used in the full algorithm for IRS in Appendix \ref{app:algorithm}.
\begin{enumerate}[leftmargin=*]
    \item \textbf{Initialise.} Set the $X$- and $Z$-sector marginals of every data qubit to zero at its reset.
    \item \textbf{Advance to the next tCNOT layer.} For each intervening syndrome-extraction round, compose into each qubit's marginals the contribution from every physical gate it participates in, using Eq. \eqref{eq:PARX}. The result is $p_{\mathrm{acc}}$ for that qubit and sector at the gate boundary.
    \item \textbf{Snapshot the layer.} Record the marginals of every data qubit before any of them are updated. The propagated contamination $p_{\mathrm{prop}}$ is read from this snapshot, so that both qubits involved in a tCNOT are advanced from the same pre-gate state.
    \item \textbf{Form the added mechanisms.} The physical tCNOT has the same two-qubit depolarizing channel as any other CNOT, so $p_{\mathrm{dir}}$ is given by Eq. \eqref{eq:pX}. For each qubit and sector receiving a contribution, set $q = \mathrm{PAR}(p_{\mathrm{prop}}, p_{\mathrm{dir}})$ and place it as an edge between the two detectors specified by the placement rule above, or to the boundary where only one supporting stabilizer is relevant. 
    \item \textbf{Merge duplicate edges.} The memory-equivalent DEM will generally already contain an edge between those detectors, since an ordinary data-qubit error in that round would trigger the same pair; the added mechanism is then merged with it by parity rather than introducing an additional parallel edge. The mechanisms obtained for block $b$ in this layer constitute $Q_{b,\ell}$.
    \item \textbf{Advance past the layer.} Replace each marginal by the corresponding $p_{\mathrm{out}}$ of Eq. \eqref{eq:fourway} and return to step 2, until all $L$ layers have been processed.
    \item \textbf{Assemble and stage.} The prior-informed DEM of block $b$ is $M_b \cup \bigcup_{\ell} Q_{b,\ell}$. During the first iteration, layer $k$ is decoded against $M_b \cup \bigcup_{\ell \geq k} Q_{b,\ell}$, so that by the end of that iteration the decoder has returned to $M_b$ and remains there for every subsequent iteration.
\end{enumerate}

Note that the marginal construction described here tracks the $X$- and $Z$-sector probabilities independently. 
A seemingly more complete variant propagates the full single-qubit Pauli distribution $\{I,X,Z,Y\}$ for every data qubit so that, at each tCNOT, the two single-qubit distributions are combined into the sixteen-element joint Pauli distribution, transformed under the exact CNOT conjugation map, composed with the two-qubit depolarizing channel, and then marginalised back onto the two constituent data qubits. 
This retains the $Y$-error contributions and the correlations between the $X$ and $Z$ sectors that the marginal approximation appears to discard. 
We have implemented this variant and find that it returns the same prior as the independent marginal construction. The reason for this is that the additional information retained by the full single-qubit Pauli distribution is the correlation between the $X$ and $Z$ Pauli sectors on the same qubit, whereas the prior-informed DEM is constructed only from their separate marginals. 
Labelling each single-qubit Pauli by its symplectic pair $(x,z)\in\mathbb{F}_2^2$, with $I=(0,0)$, $X=(1,0)$, $Z=(0,1)$ and $Y=(1,1)$, Pauli composition is addition modulo two on the pair. 
The quantities entering the DEM are therefore only the marginal probabilities $\Pr[x=1]$ and $\Pr[z=1]$ for each data qubit. 
Composition with an independent Pauli channel preserves this separation: the channel's $x$ bit is added only to the existing $x$ bit, and its $z$ bit only to the existing $z$ bit. 
Equivalently, in the sign representation $1-2\Pr[x=1]$, independent $z$-type contributions multiply among themselves, as do $x$-type contributions. 
For a CNOT, let $(c_x,c_z)$ and $(t_x,t_z)$ denote the symplectic bits of the Pauli errors on the control and target immediately before the gate, and let primes denote the corresponding bits after conjugation. 
CNOT conjugation gives
\begin{align}
    c_x'=c_x,\quad c_z'=c_z\oplus t_z, \nonumber \\
    t_x'=c_x\oplus t_x,\quad t_z'=t_z,
\end{align}
so every output $x$ bit depends only on input $x$ bits, while every output $z$ bit depends only on input $z$ bits. 
Neither operation used to construct the prior mixes the two sectors, and the within-qubit correlation between $x$ and $z$ retained by the full Pauli distribution is never required when computing the marginal probabilities supplied to the DEM. 
The $X$- and $Z$-sectors consequently evolve as two decoupled $\mathbb{F}_2$-linear systems; this explains why propagating the full $\{I,X,Z,Y\}$ distribution gives the same prior as propagating the two marginals independently. 
Across all circuit families tested, we find that the two constructions agree to $1.5\times10^{-12}$, consistent with floating-point precision rather than any residual approximation error.

\section{Additional GHZ decoding results}\label{app:ghz-decoding}
Figure~\ref{fig:ghz-additional} supplements Figure~\ref{fig:quera_ghz_graph} with a comparison of correlated and uncorrelated matching (correlated pymatching within IRS), and with LOM~\cite{serra2026decoding} which only uses uncorrelated pymatching in its open source implementation. 
With the reference half-full initialisation schedule, LOM agrees within statistical uncertainty with reliable logical Pauli product (RLPP) decoding~\cite{cain2025fast}. 
The two methods share the same observable-based matching principle, and this numerical agreement provides a consistency check on our benchmark implementation. 
With full-full initialisation, LOM gives lower logical error rates than uncorrelated IRS but higher rates than correlated IRS. 
The choice of local matching therefore makes a substantial contribution to the IRS advantage on this benchmark.
\begin{columnfigure}[htbp]
    \centering
    \includegraphics[width=\linewidth]{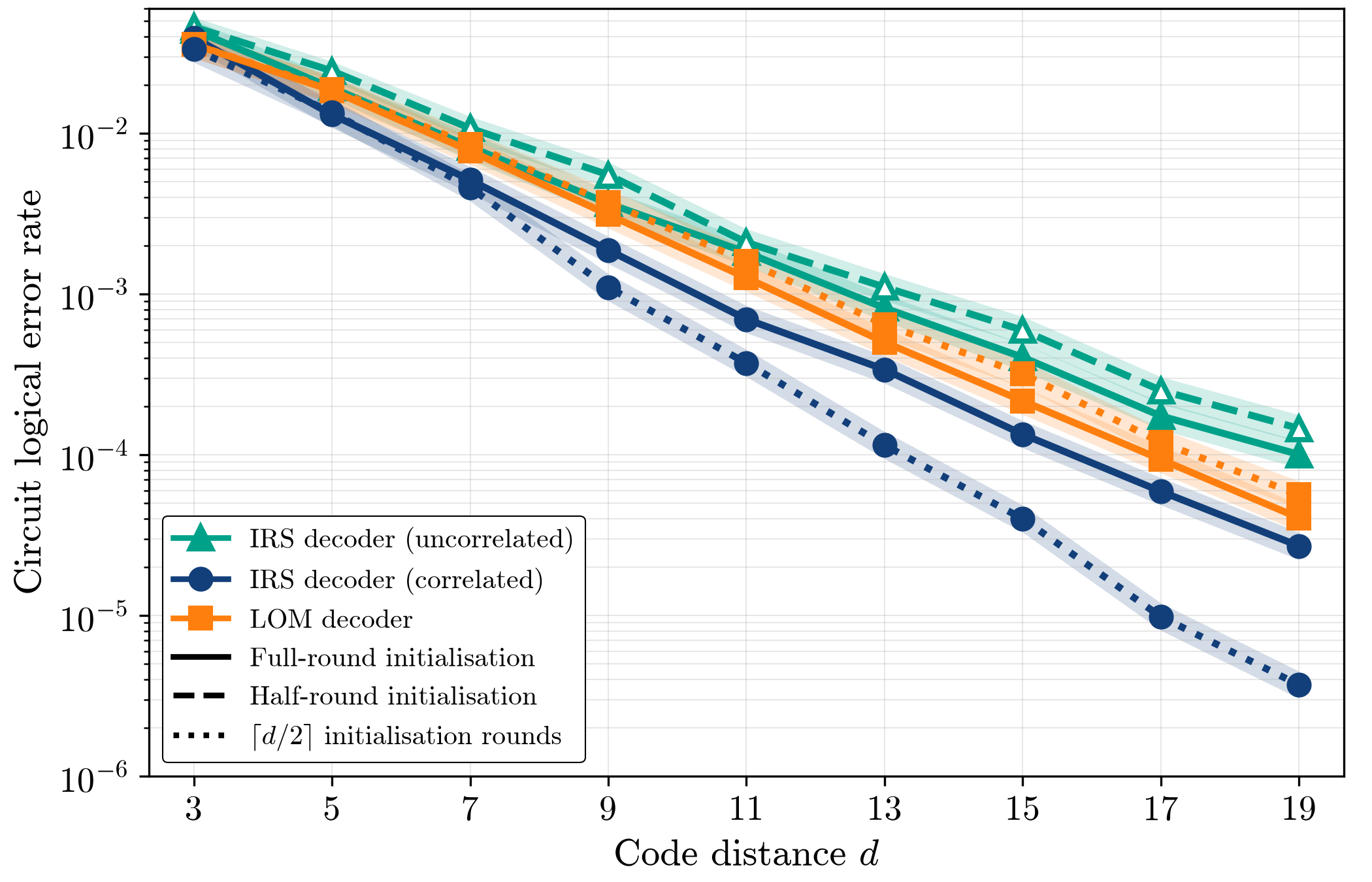}
    \caption{Additional comparisons for the $3$-block GHZ benchmark of Figure~\ref{fig:quera_ghz_graph}, using the logical circuit and physical error model of Cain et al.~\cite{cain2025fast}. Teal triangles and blue circles show IRS with uncorrelated and correlated local MWPM, respectively. Orange squares show results using the LOM decoder of Serra-Peralta et al.~\cite{serra2026decoding}. Solid and dashed lines use full-full and half-full round initialisation schedules, respectively, while the dotted line uses $\lceil d/2\rceil$ full syndrome-extraction rounds before the first tCNOT. Shaded bands indicate the 95\% confidence Wilson interval.}
    \label{fig:ghz-additional}
\end{columnfigure}

Increasing the initialisation to $\lceil d/2\rceil$ full rounds gives stronger distance-dependent suppression for correlated IRS, but no corresponding improvement for LOM, whose error rate instead increases slightly at larger distances. At $d=19$, correlated IRS gives a logical error rate approximately fifteen times lower than LOM with this longer schedule. This supports the distinction in Section~\ref{sec:initial-stabilizers}: completing the initial half round can benefit both approaches, whereas adding further full rounds provides an additional improvement only for IRS in these tests.

We further test a single-tCNOT Bell-state preparation with $XZ$ initialisation, where initially indeterminate stabilizers interact nontrivially with the IRS decoding problem, against a $ZZ$-initialised reference circuit. Figure~\ref{fig:bell-init-rounds} shows that $\lceil d/2\rceil$ initial full syndrome-extraction rounds are sufficient to recover the reference logical error rate at all distances tested, with no further improvement from additional rounds.

\begin{columnfigure}[htbp]
    \centering
    \includegraphics[width=\linewidth]{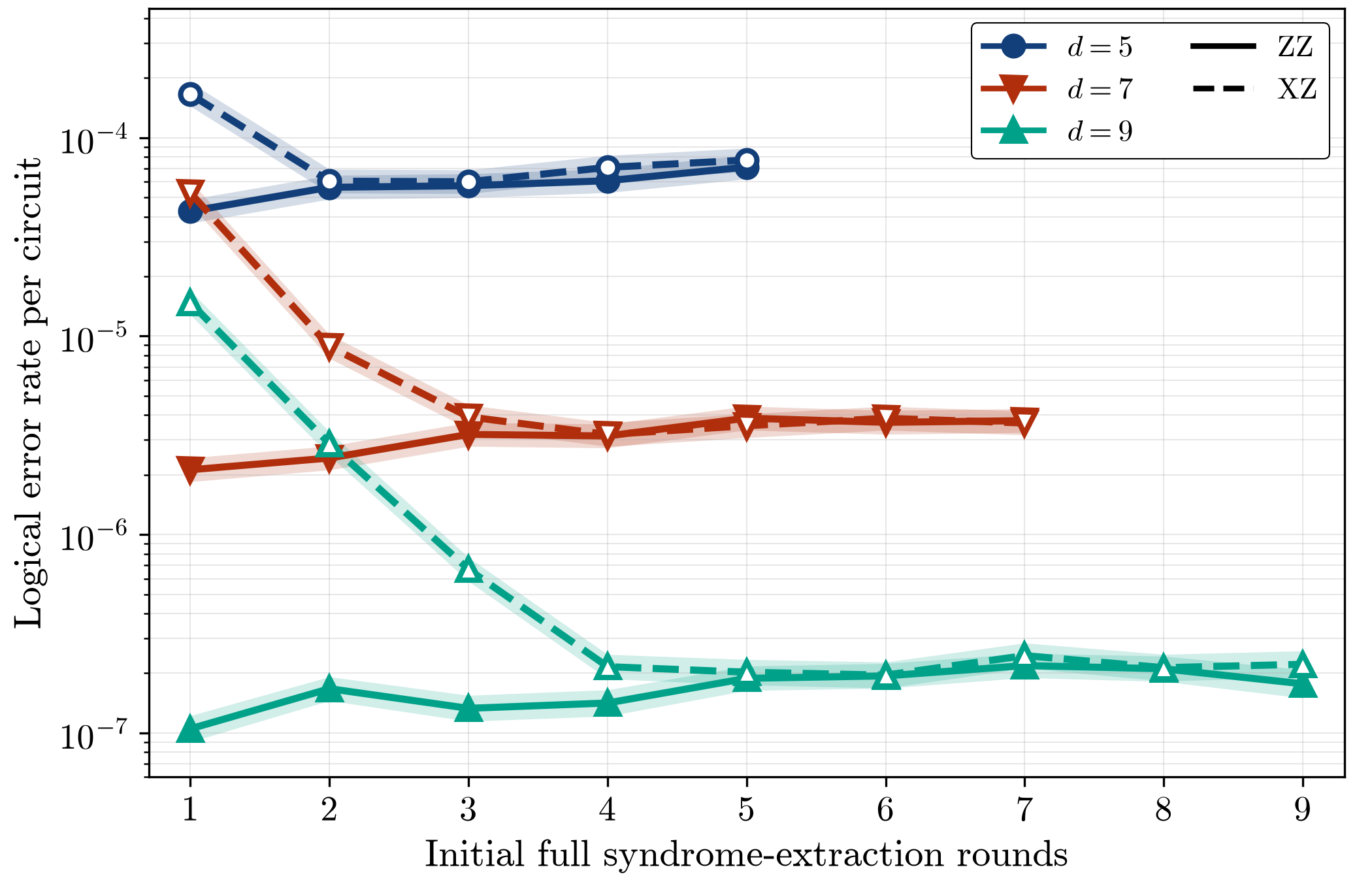}
    \caption{Logical error rates for single-tCNOT circuits under IRS, comparing Bell-state preparation ($\ket{+}_L\ket{0}_L$, dashed line) against a non-entangling reference ($\ket{0}_L\ket{0}_L$, solid line) at $d=5,7,9$. The initialisation-induced decoding penalty is removed by approximately $\lceil d/2\rceil$ full syndrome-extraction rounds.}
    \label{fig:bell-init-rounds}
\end{columnfigure}

\section{Circuit Details}
\subsection{Error models}\label{app:noise-models}
We use two circuit-level Pauli noise models, with faults sampled independently at each location. Depolarizing probabilities denote the total probability of a nonidentity Pauli, distributed uniformly over the three single-qubit or fifteen two-qubit Paulis. IRS uses correlated PyMatching, with the exception of Figure \ref{fig:bell-init-rounds}, where both correlated and uncorrelated decoding are compared. The Stim circuits and numerical data for all benchmarks are available in the accompanying Zenodo repository~\cite{webber2026stim}.

\subsubsection{Default model}
The deep Clifford benchmark (Figure~\ref{fig:deep-random}), alternating-tCNOT tests (Figures~\ref{fig:iteration_dependence} and \ref{fig:calibration}), and repeated-$S$ circuits (Figure~\ref{fig:S-gate-decoding}) use a model matched to Turner et al.~\cite{turner2026scalable} and their open-source Stim circuits, with $p=10^{-3}$. 
Physical CNOT and $CZ$ gates are followed by two-qubit depolarization with probability $p$, including those implementing logical gates. 
Preparation/reset and the physical single-qubit gates implementing logical operations are followed by single-qubit depolarization with probability $p/10$. 
The same channel acts on inactive qubits during each syndrome-extraction CNOT substep. 
Ancilla measurement outcomes flip with probability $p$. 
The deep Clifford, alternating-tCNOT and repeated-$S$ benchmarks terminate with ideal MPP measurements of stabilizers and logical observables. 
The GHZ benchmarks instead use destructive single-qubit $Z$ readout with outcome-flip probability $p$, as detailed below.

The repeated-$S$ gadget uses separate $CZ$ and $S$ substeps, with the corresponding gate noise and $p/10$ depolarization on idle data qubits in each substep. 
The alternating-tCNOT memory references retain the extraction schedule but remove the logical CNOTs and their gate noise.

\subsubsection{GHZ model}
The GHZ results in Figures~\ref{fig:quera_ghz_graph} and \ref{fig:ghz-additional} use the circuit and error model of Cain et al.~\cite{cain2025fast}, supplied directly by the authors in private communication, with $p=3\times10^{-3}$. 
All physical CNOTs are followed by two-qubit depolarization with probability $p$. 
Preparation/reset and idle locations have single-qubit depolarization with probability $p$, rather than $p/10$. 
Both ancilla measurements and final destructive data measurements have outcome-flip probability $p$. 
We retain the supplied circuit's additional data-only waiting layers, including four independent idle channels on each spectator data qubit while the other two patches undergo syndrome extraction.

\subsection{Circuit construction}\label{app:circuit-construction}
The Stim circuits used for sampling, the memory-equivalent circuits used to construct the IRS decoding DEMs, and the numerical data underlying the paper's figures are available in the accompanying Zenodo repository~\cite{webber2026stim}.

The deep Clifford benchmark of Figure \ref{fig:deep-random} uses a fixed sequence of Hadamard, Pauli and tCNOt, first used in Ref.~\cite{cain2024correlated}, and later used as a benchmark in Ref~\cite{turner2026scalable}; the authors of the latter provided us with the exact circuit.
A logical gate layer consists of randomly chosen single-qubit transversal gates from $\{H, X, Y, Z\}$ on each logical qubit, followed by syndrome-extraction buffer rounds and then two transversal CNOTs acting on randomly chosen disjoint pairs of logical qubits.
Each logical qubit is prepared in $\ket{+}_L$, followed by two initial syndrome-extraction rounds and 32 logical gate layers as defined above. A final syndrome-extraction buffer precedes perfect measurements of the logical stabilizers of the state.

Alternating-tCNOT circuits repeatedly reverse the control and target roles of two blocks.
The circuits used in Figure \ref{fig:iteration_dependence} are for chains of alternating tCNOTs $CX_{A\rightarrow B}CX_{B\rightarrow A}CX_{A\rightarrow B}CX_{B\rightarrow A}\ldots$ with a fixed SE round schedule, whereas Figure \ref{fig:calibration} is for a single $CX_{A\rightarrow B}CX_{B\rightarrow A}$ pair with a range of SE round buffers.
The black 'memory-equivalent' references data in Figure \ref{fig:calibration} retain the same SE schedule as the other circuits, but with both logical CNOTs and their associated two-qubit depolarization removed. 
For both figures, logical qubits are initialised in $\ket{0}_L$, with each block being measured individually in the $Z$-basis.

The GHZ circuit in Figure \ref{fig:quera_ghz_graph} prepares $\ket{+}_L\ket{0}_L\ket{0}_L$, applies tCNOTs $0\to1$ and $0\to2$, and measures $Z_L^{(1)}Z_L^{(2)}$.

The repeated-$S$ circuit in Figure \ref{fig:S-gate-decoding} prepares $\ket{+}_L$, applies $d+1$ logical $S$ gates separated by one full extraction round, and measures $X_L$. 
Initialisation and final extraction-round counts follow the schedules specified in the figure.

All benchmark Stim circuits, together with separate Stim circuits used to generate the decoding DEMs, are available in the accompanying Zenodo archive.

\end{appendices}

\bibliography{main}
\end{document}

%% file: preamble.tex
\PassOptionsToPackage{normalem}{ulem}

\usepackage[utf8]{inputenc}
\usepackage[T1]{fontenc}
\usepackage{babel}
\usepackage{libertine}

\usepackage[letterpaper]{geometry}
\usepackage{setspace}
\usepackage{enumitem}
\usepackage{ulem}
\usepackage{microtype}

\usepackage{titlesec}
\titlespacing{\section}{0pt}{14pt}{8pt}
\titlespacing{\subsection}{0pt}{10pt}{6pt}
\titlespacing{\subsubsection}{0pt}{8pt}{5pt}

\usepackage[title]{appendix}

\usepackage[dvipsnames]{xcolor}
\usepackage{xreview}
\usepackage{tcolorbox}

\usepackage{mathtools}   
\usepackage{amssymb}
\usepackage{wasysym}
\usepackage{bm}
\usepackage{bbm}
\usepackage{bbold}

\usepackage{graphicx}
\usepackage[caption=false]{subfig}
\usepackage{svg}

\usepackage{float}

\makeatletter
\let\newfloat\newfloat@ltx
\makeatother

\usepackage{algorithm}
\usepackage{algpseudocode}

\usepackage{tikz}

\usetikzlibrary{
    decorations.pathreplacing,
    quantikz2,
    shapes.geometric,
    arrows.meta,
    positioning,
    calc
}

\usepackage{orcidlink}
\usepackage{xspace}
\usepackage{url}
\usepackage[scaled=0.90]{zi4}       

\renewcommand{\thesection}{\arabic{section}}
\renewcommand{\thesubsection}{\thesection\alph{subsection}}

\tikzset{
    startstop/.style={
        rectangle,
        rounded corners=8pt,
        draw=black,
        thick,
        minimum width=3.5cm,
        minimum height=0.8cm,
        text centered,
        font=\small
    },
    process/.style={
        rectangle,
        draw=black,
        thick,
        minimum width=5cm,
        minimum height=0.8cm,
        text centered,
        font=\small
    },
    output/.style={
        trapezium,
        trapezium left angle=70,
        trapezium right angle=110,
        draw=black,
        thick,
        minimum width=4.5cm,
        minimum height=0.8cm,
        text centered,
        font=\small
    },
    decision/.style={
        diamond,
        aspect=2.5,
        draw=black,
        thick,
        minimum width=5cm,
        minimum height=1cm,
        text centered,
        font=\small
    },
    combine/.style={
        rectangle,
        draw=black,
        thick,
        minimum width=5.5cm,
        minimum height=0.8cm,
        text centered,
        font=\small
    },
    arrow/.style={
        -{Stealth[length=6pt]},
        thick
    },
    dasharrow/.style={
        -{Stealth[length=6pt]},
        thick,
        dashed
    }
}

\DeclareTextSymbolDefault{\textquotedbl}{T1}

\DeclareFontFamily{OT1}{pzc}{}
\DeclareFontShape{OT1}{pzc}{m}{it}
    {<-> s * [1.05] pzcmi7t}{}
\DeclareMathAlphabet{\mathpzc}{OT1}{pzc}{m}{it}

\definecolor{Review}{RGB}{200,30,30}

\algnewcommand{\To}{\textbf{To }}
\algnewcommand\Input{\item[\textbf{Input:}]}
\algnewcommand\Output{\item[\textbf{Output:}]}

\newenvironment{columnfigure}[1][tbp]
{%
    \begin{figure}[#1]%
    \centering
}
{%
    \end{figure}%
}

\ifTwoColumn
    \newenvironment{widefigure}[1][t]
    {%
        \begin{figure*}[#1]%
        \centering
    }
    {%
        \end{figure*}%
    }
\else
    \newenvironment{widefigure}[1][t]
    {%
        \begin{figure}[#1]%
        \centering
    }
    {%
        \end{figure}%
    }
\fi